\documentclass[10pt,aps,prd,reprint,nofootinbib,superscriptaddress,preprintnumbers]{revtex4-2}

\usepackage{graphicx}
\usepackage{amssymb}
\usepackage{amsmath}
\usepackage{fontawesome}
\usepackage{xcolor}
\usepackage{bbm}
\usepackage{multirow}
\usepackage[colorlinks,pdfusetitle,breaklinks,linktocpage=true]{hyperref}
\hypersetup{colorlinks=true,allcolors=[rgb]{1,0.56,0}}

\newcommand{\orcid}[1]{\href{https://orcid.org/#1}{#1}}
\newcommand{\e}[1]{\times10^{#1}}

\usepackage{listings}
\definecolor{dkgreen}{rgb}{0,0.6,0}
\definecolor{gray}{rgb}{0.5,0.5,0.5}
\definecolor{mauve}{rgb}{0.58,0,0.82}
\definecolor{codegreen}{rgb}{0,0.6,0}
\definecolor{codegray}{rgb}{0.5,0.5,0.5}
\definecolor{codepurple}{rgb}{0.58,0,0.82}
\definecolor{backcolour}{rgb}{0.95,0.95,0.92}
\lstdefinestyle{mystyle}{
    backgroundcolor=\color{backcolour},   
    commentstyle=\color{codegreen},
    keywordstyle=\color{magenta},
    numberstyle=\tiny\color{codegray},
    stringstyle=\color{codepurple},
    basicstyle=\ttfamily\footnotesize,
    breakatwhitespace=false, 
    breaklines=true, 
    captionpos=b,    
    keepspaces=true, 
    numbers=none,
    numbersep=5pt,
    showspaces=false,
    showstringspaces=false,
    showtabs=false,  
    tabsize=3,
}
\begin{document}

\preprint{FERMILAB-PUB-25-0704-T}

\title{\texttt{NuFast-Earth}: Efficient Atmospheric, Solar, and Supernova Neutrino Propagation Through the Earth}

\author{Peter B.~Denton}
\email{pdenton@bnl.gov}
\thanks{\orcid{0000-0002-5209-872X}}
\affiliation{High Energy Theory Group, Physics Department \\ Brookhaven National Laboratory, Upton, NY 11973, USA}

\author{Stephen J.~Parke}
\email{parke@fnal.gov}
\thanks{\orcid{0000-0003-2028-6782}}
\affiliation{Theoretical Physics Department, Fermi National Accelerator Laboratory, Batavia, IL 60510, USA}

\date{March 10, 2026}

\begin{abstract}
Algorithms for computing neutrino oscillation probabilities in sharply varying matter potentials such as the Earth are becoming increasingly important.
As the next generation of experiments, DUNE and HyperK as well as the IceCube upgrade and KM3NeT, come online, the computational cost for atmospheric and solar neutrinos will continue to increase.
To address these issues, we expand upon our previous algorithm for long-baseline calculations to efficiently handle probabilities through the Earth for atmospheric, nighttime solar, and supernova neutrinos.
The algorithm is fast, flexible, and accurate.
It can handle arbitrary Earth models with two different schemes for varying density profiles.
We also provide a \texttt{c++} implementation of the code called \texttt{NuFast-Earth} along with a detailed user manual.
The code intelligently keeps track of repeated calculations and only recalculates what is needed on each successive call which can also help provide significant speed-ups.
\begin{center}
\href{https://github.com/PeterDenton/NuFast-Earth}{\large\faGithub}
\end{center}
\end{abstract}

\maketitle

\tableofcontents

\section{Introduction}
The Wolfenstein matter effect \cite{Wolfenstein:1977ue} provides a key modification to neutrino oscillation probabilities in many of the environments where we observe neutrino oscillations.
The matter effect can muddy the extraction of some fundamental oscillation parameters and yet is also essential in the extraction of others, see e.g.~\cite{Denton:2025jkt}.
Nonetheless, other than solar neutrinos experiencing the effect inside the Sun, there has not yet been a clear detection of the matter effect in other environments, notably within the Earth.

Current and next-generation experiments are aiming to make incredibly precise measurements of neutrino oscillations from a variety of sources.
Atmospheric neutrinos will provide precision measurements of $\sin^2 2\theta_{23}$ and $\Delta m^2_{32}$ as well as some information on the octant of $\theta_{23}$, $\theta_{13}$, the mass ordering, and $\delta$ \cite{Kelly:2019itm,Martinez-Soler:2019nhb,Arguelles:2022hrt,Suliga:2023pve} and solar neutrinos should provide \cite{DUNE:2024wvj,Hyper-Kamiokande:2018ofw,Capozzi:2018dat} the first clear measurement of the day-night effect \cite{Baltz:1986hn,Bouchez:1986kb,Carlson:1986ui,Cribier:1986ak} which the current data is only beginning to hint at \cite{Super-Kamiokande:2023jbt}.
Extracting the most information possible from these measurements requires incredibly precise predictions for the event rates which, in practice, requires many computational throws, using Feldman-Cousins \cite{Feldman:1997qc} techniques, see e.g.~\cite{nova_computing:2018}.
Each such throw requires recomputing the oscillation probability which is often the slowest aspect of the computation, even in long-baseline accelerator neutrinos where only one distance needs to be considered \cite{novaprivate}.

To address this issue, for long-baseline accelerator and reactor neutrino oscillations, we developed the \texttt{NuFast-LBL} algorithm \cite{Denton:2024pzc} and the associated code on github \cite{nufast-lbl-github}.
This algorithm was based on a variety of theoretical neutrino oscillation results obtained over a number of years including some of the earliest full expressions for neutrino oscillations in constant matter \cite{Barger:1980tf,Zaglauer:1988gz}, solutions to the cubic equation \cite{cardano,Denton:2024pzc}, the role of CP violation \cite{Jarlskog:1985ht,Bilenky:1987ty,Krastev:1988yu,Naumov:1991ju,Harrison:2002ee,Denton:2019yiw,Parke:2020wha}, the optimal means of computing the eigenvectors \cite{Denton:2018fex,Denton:2019ovn,Denton:2019pka,Denton:2019qzn,Abdullahi:2022fkh,Denton:2024pzc,Kopp:2006wp}, and the optimal structure for the oscillation probabilities as well as various other theoretical advances \cite{Barenboim:2019pfp,Agarwalla:2013tza,Minakata:2015gra,Denton:2016wmg,Denton:2018hal},
see also \cite{Ohlsson:1999xb,Huber:2004ka,Nunokawa:2005nx,Huber:2007ji,Parke:2016joa,Denton:2018cpu,Wang:2019yfp,Arguelles:2021twb,Maltoni:2023cpv,Luo:2023xmv,Page:2023rpb,Gonzalo:2023mdh,Denton:2024thm}.

In this paper, we use parts of the \texttt{NuFast-LBL} algorithm and modify it to handle the varying density profiles through the Earth in a novel calculation.
The additional complexities of traversing sharply varying density profiles require a new approach constructed from the ground up, which is the result presented in this paper.
We provide a new code called \texttt{NuFast-Earth}, also available on github, see \cite{nufast-earth-github}\footnote{The code is available at \href{https://github.com/PeterDenton/NuFast-Earth}{github.com/PeterDenton/NuFast-Earth}.}, which is significantly more powerful than the long-baseline (LBL) code.
Notably, it intelligently saves reused aspects of the calculations for future uses which vastly speeds up many calculations of the oscillation probabilities.
It is also designed to handle arbitrary spherical Earth density models with numerous models coded up and the ability to easily add more.

In this paper, we describe how we implement the Earth in section \ref{sec:earth}.
Subsection \ref{sec:eigenvalues and eigenvectors} discusses the techniques used to compute the eigenvalues and the eigenvectors, specifically what we call the internal eigenvectors, with calculational details given in appendix \ref{app:Adj}.
We then discuss how we take the internal amplitude constructed from those eigenvalues and eigenvectors and combine it into a probability in subsection \ref{sec:amplitudes}.
This completes the calculation for atmospheric neutrinos.
The solar and supernova calculations require several additional steps discussed in section \ref{sec:astro}.
Numerical results for a variety of cases are shown in section \ref{sec:results} with additional numerical results shown in appendix \ref{sec:validation}.
We briefly describe the code usage in section \ref{sec:code}; see also the more detailed user's guide on the github page.
In section \ref{sec:speed} we perform detailed speed and precision tests and make recommendations on how to optimally use the code depending on one's usage case and we conclude in section \ref{sec:conclusions}.

\section{Earth}
\label{sec:earth}
This paper is focused on the problem of neutrinos propagating through the Earth.
Neutrinos propagating in other dense environments such as the Sun, supernova, the early Universe, and other celestial bodies each have their own unique challenges depending on particle content, density, typical neutrino energies, geometry, and temporal variation.
For this discussion, the matter density of protons, neutrons, and electrons in the Earth are significant enough to appreciably modify neutrino oscillations, but not so large as to lead to absorption, which does not become relevant until $E\gtrsim\mathcal O($TeV$)$, at which point there are oscillations are extremely damped in the standard three-flavor picture.

\subsection{Earth Models}
The majority of the information we have about the interior of the Earth comes from seismic information.
In the past, models of the Earth's density profile assume spherical symmetry, but recently there has been some evidence of some small non-symmetric substructures \cite{doi:10.1080/08120090801888578,https://doi.org/10.1029/GL013i013p01545,https://doi.org/10.1029/94GL01600,tanaka2007possibility,song1998seismic,karato2000earth,ishii2002innermost,frost2021dynamic}.
Nonetheless, we will focus on spherically symmetric Earth models as this leads to some computational advantages and the expected future sensitivity to such substructures in the Earth with atmospheric \cite{Nicolaidis:1990jm,Winter:2006vg,Akhmedov:2006hb,Agarwalla:2012uj,Rott:2015kwa,Winter:2015zwx,DOlivo:2020ssf,Kumar:2021faw,Kelly:2021jfs,Denton:2021rgt,Capozzi:2021hkl,DOlivoSaez:2022vdl,Maderer:2022toi,Upadhyay:2022jfd,Raikwal:2023jkf,Upadhyay:2024gra,Petcov:2024icq,Jesus-Valls:2024tgd,Chattopadhyay:2025cgt,J:2025kdq,Chattopadhyay:2025ulr}, solar \cite{Akhmedov:2005yt,Bakhti:2020tcj}, or supernova \cite{Lindner:2002wm,Akhmedov:2005yt,Hajjar:2023knk} neutrino oscillations is weak.
A more complicated asymmetric Earth model could be implemented in the code via the single trajectory mode, but is not the focus of this paper.

We consider the Earth's density profile to be smoothly varying except at a finite number of sharp discontinuities.
We base our density models on the Preliminary Reference Earth Model (PREM) \cite{Dziewonski:1981xy} and use the electron fractions from \cite{Mocioiu:2000st}.
We assume the atmosphere to be true vacuum\footnote{This is an excellent approximation.
We have calculated the maximum shift in the probabilities across $E\in[2,40]$ GeV and $\cos\theta_z\in[-1,0.1]$ to be:
\begin{center}
\begin{tabular}{|c|c|c|c|c|}
\hline
\multirow{2}{*}{$\max|\Delta P|$} & \multicolumn{2}{c|}{$\nu_\mu\to\nu_\mu$} & \multicolumn{2}{c|}{$\nu_\mu\to\nu_e$}\\\cline{2-5}
& $\nu$ & $\bar\nu$ & $\nu$ & $\bar\nu$ \\\hline
NO & $5\e{-8}$ & $3\e{-8}$ & $3\e{-7}$ & $3\e{-7}$\\\hline
IO & $8\e{-8}$ & $4\e{-8}$ & $3\e{-7}$ & $3\e{-7}$\\\hline
\end{tabular}
\end{center}
where the production height is 10 km, the detector depth is 2 km, and the atmosphere is either true vacuum or has $\rho Y_e=0.5\times10^{-3}$ g/cm$^3$ which is conservative, e.g.~large.}.

We also provide various modifications to the Earth model in the code.
In one, we merge the top several layers of the PREM together so that there are only four layers in total.
In another, we take the four constant layers provided in the \texttt{prob3++} code \cite{Barger:1980tf}.
We also provide two other very useful Earth models.
In one we take $N$ layers of constant density sampled from the PREM distribution, and in the other we set the density of the entire Earth as constant. Finally, it is simple for the user to insert their own Earth model of arbitrary spherical complexity.
More details on the prebuilt Earth models are provided in the user's guide.

\subsection{Trajectories}
\label{subsec:Traj}
As the neutrinos pass from the production point, through the atmosphere, into the Earth, across the Earth and to the detector, the densities that they experience vary considerably.
We break the trajectory up into four main segments.
First, we start from the production point in the atmosphere to the surface of the Earth.
As this section is purely in vacuum it is very easy to compute and modify.
Second, we go from the surface of the Earth to the detector depth.
Third, we go from the detector depth to the innermost point in the Earth.
Finally, we go from the innermost point in the Earth to the detector depth at the actual detector which is the same as the previous step, up to a transpose for a spherically symmetric Earth.
Reusing the calculations for one of these segments from the other saves a sizable amount of computational effort.
These trajectories are shown in Fig.~\ref{fig:Traj_Sch} and are discussed in more detail below.

\section{The Algorithm}
In this section, we describe the algorithm used to compute neutrino oscillation probabilities.
We start with the oscillation amplitude in a single constant density layer and then combine these amplitudes to form the full amplitude for a given neutrino trajectory. Finally, we extract the oscillation probabilities.
\subsection{One Layer Amplitude}
\label{sec:eigenvalues and eigenvectors}
Unlike in the LBL case where we work with the squared amplitude in the probability form, for atmospheric neutrinos we need to work directly with the amplitude.
We first compute the amplitude through a single layer of approximately constant density.
The oscillation amplitude for a given layer with a chord length, $L$, in the layer and neutrino energy, $E$, is given by the unitary matrix
\begin{align}
    {\cal A}_{\alpha \beta} = \delta_{\alpha \beta} &+V_{\alpha 2}V^*_{\beta 2}(e^{-i \Delta \lambda_{21}L/(2E)}-1) \notag \\
    &+V_{\alpha 3}V^*_{\beta 3}(e^{-i \Delta \lambda_{31}L/(2E)}-1)\,,
    \label{eq:amp-layer}
\end{align}
where $\Delta\lambda_{ij}=\lambda_i-\lambda_j$, the $\lambda_i$'s are the solutions to the characteristic equation (and are related to the $\Delta m^2$'s in matter), and $V_{\alpha i}$ is the $\alpha$ element of the eigenvector (which is related to the mixing angles in matter) for $\lambda_i$ of the Hamiltonian of the given layer.
In vacuum $\Delta \lambda_{ij} = \Delta m^2_{ij}$ and $V_{\alpha i} = U_{\alpha i}$, the PMNS mixing matrix.
This form of the amplitude is obtained from the standard expression,
\begin{equation}
\sum_i V_{\alpha i}V^*_{\beta i}e^{-i \lambda_{i}L/(2E)}\,,
\end{equation}
by removing an overall phase and using the unitarity of $V$ to derive eq.~\ref{eq:amp-layer}, which requires fewer costly calls to the trigonometric or exponential functions.

The PMNS matrix that is used is given by
\begin{align}
    U &\equiv R_{23}(\theta_{23}) D_{3}(\delta) R_{13}(\theta_{13}) R_{12}(\theta_{12})\,,
\end{align}
where 
\begin{align}
    R_{23}(\theta_{23}) &\equiv 
    \left( \begin{array}{ccc}
        1 &&\\
         &c_{23} & s_{23}\\
         &-s_{23} & c_{23}
    \end{array}
    \right), \quad
    \notag \\
    R_{13}(\theta_{13}) &\equiv 
    \left( \begin{array}{ccc}
        c_{13} && s_{13}\\
         &1 & \\
         -s_{13} & &c_{13}
    \end{array}
    \right)\,, \notag \\
    R_{12}(\theta_{12}) &\equiv 
    \left( \begin{array}{ccc}
        c_{12} & s_{12}&\\        
         -s_{12} & c_{12}&\\
         &&1
    \end{array}
    \right)\,, \notag \\
    D_3(\delta) &\equiv \text{Diag}(1,1,\,e^{i\delta}\, )\,,
        \end{align}
using the shorthand $s_{ij}=\sin \theta_{ij}$ and $c_{ij}=\cos \theta_{ij}$.  
This choice is equivalent to the PDG choice \cite{ParticleDataGroup:2024cfk}, since if one right multiplies by
$D_3(-\delta)$, one obtains the PDG PMNS matrix, see, e.g.,~\cite{Denton:2020igp}.
As this is a change of the phase of $\nu_3$, 
no oscillation physics observable can be sensitive to this phase change.
As will be shown shortly, 
we have a specific computational reason for making the above choice for the PMNS matrix.

The Hamiltonian in the flavor basis is given by
\begin{align}
    (2E)H_f = UM^2U^\dagger + W\,,
    \label{eq:HF}
\end{align}
where $M^2 = \text{Diag}(0, \Delta m^2_{21}, \Delta m^2_{31})$ and the Wolfenstein matter potential is given by $W \equiv \text{Diag}(A_{mat},0,0)$ with $A_{mat} \equiv  \pm 2\sqrt{2}\, G_F N_e E$ (+ for neutrinos and - for antineutrinos; the sign of $\delta$ also flips between neutrinos and antineutrinos).
This Hamiltonian is a Hermitian matrix with complex entries because the $D_3(\delta)$ rotation.
Since $R_{23}(\theta_{23})D_3(\delta)$ commutes with the matter potential, we separate out the $R_{23}(\theta_{23})D_3(\delta)$ rotation for the majority of the calculation, and then apply it at the very end of the full amplitude calculation.
This is the maximal amount of information that can be rotated out of the Hamiltonian.

The Hamiltonian that we use for a significant part of the calculation is 
\begin{align}
    (2E)\widetilde H = R_{13} R_{12}M^2 R^T_{12}R^T_{13} + W\,,
    \label{eq:H}
\end{align}
which is a real symmetric matrix that not only has real eigenvalues, the $\lambda$'s, but the eigenvectors $\widetilde V_{\alpha i}$ can also be chosen to be all real\footnote{We have suppressed the arguments of the rotation matrices for readability.}.
We call this basis the ``tilde'' basis, see e.g.~\cite{Denton:2016wmg}, which is related to the flavor basis by
\begin{equation}
 H_f=R_{23}D_3 \widetilde HD_3^*R_{23}^T\,,
\label{eq:tildeH to Hf}
\end{equation}
where $D_3^*(\delta)=D_3(-\delta)$.
With this choice the oscillation amplitude in one layer, eq.~\ref{eq:amp-layer}, is a complex symmetric matrix, therefore only 6 of the 9 elements need to be explicitly calculated.

Progressing for now in the tilde basis, the eigenvalues of $(2E)\widetilde H$, eq.~\ref{eq:H}, are given by the solutions to the characteristic equation,
\begin{equation}
\text{Det}[\lambda I -(2E)\widetilde H]=0\,,
\end{equation}
which are all real and the relevant components of the normalized eigenvectors are calculated using\footnote{The symbol ``Adj[X]'' stands for Adjugate of the matrix X, i.e. the transpose of the cofactor matrix of X.}
\begin{equation}
\widetilde V_{\alpha i}\widetilde V^*_{\beta i} = \frac{\text{Adj}[\lambda_i I -(2E)\widetilde H]}{\Pi_{k\neq i}(\lambda_i-\lambda_k)}\,.
\end{equation}
Since $\widetilde H$ is a real, symmetric matrix, all the components of all the eigenvectors can be rephased to be real, so that with this rephasing choice $\widetilde V_{\alpha i}\widetilde V^*_{\beta i}$ becomes $\widetilde V_{\alpha i}\widetilde V_{\beta i}$. 
Thus, $(2E)\widetilde H$ of eq.~\ref{eq:H} is diagonalized by a real $\widetilde V$, as follows:
\begin{align}
\widetilde V \Lambda\widetilde V^T = (2E) \widetilde H\,,
\end{align}
with $\Lambda \equiv \text{Diag}(\lambda_1, \lambda_2, \lambda_3)$.
Note that the eigenvalues of $H$ and $\widetilde H$ are the same since W commutes with $R_{23} D_3$ as noted earlier.
Further details of this diagonalization method can be found in appendix \ref{app:Adj}.

\subsubsection{Calculation of Eigenvalues}
The characteristic equation the Hamiltonian of eq.~\ref{eq:H} is given as
\begin{align}
  \chi(\lambda) &\equiv   \lambda^3 - A \lambda^2 + B \lambda -C =0\,,
  \label{eq:CharX}
\end{align}
where
\begin{align}
    A &= \Delta m^2_{31}+ \Delta m^2_{21}+A_{mat} \notag \\
    B &= \Delta m^2_{31} \Delta m^2_{21} \notag \\
     & \quad+ A_{mat} (\Delta m^2_{31} c^2_{13}+ \Delta m^2_{21}(1-c^2_{13} s^2_{12})) \notag \\
     C &= A_{mat} \Delta m^2_{31} \Delta m^2_{21} c^2_{13} c^2_{12}\,.
     \label{eq:ABC}
\end{align}
We provide two separate ways to find the solutions to this equation, one exact method by Cardano from 1545 and another that starts with an excellent approximation and then is improved to arbitrary precision by using Newton-Raphson (NR) iterations. With two iterations, this method has the same accuracy as the Cardano method using double precision arithmetic with a modest saving in computer resources.
We begin by computing $\lambda_3$ either exactly or approximately and then using $\lambda_3$ to determine $\lambda_1$ and $\lambda_2$.

\paragraph*{\bf Exact method:}
The exact analytic expression for $\lambda_{3}$ is
\begin{align}
\lambda_{3}={}&\frac1{3}A+\frac2{3} \sqrt{A^2-3B}  \\
& \times \cos\left[\frac1{3}\left(\arccos\left[\frac{2A^3-9AB+27C}{2 \left( \sqrt{A^2-3B}\,\right)^3}\right]+2\pi n \right) \right]  \notag
\end{align}
with $n=0$ for NO and $n=1$ for IO \cite{cardano,Barger:1980tf, Zaglauer:1988gz}.
As one can see this requires two costly trigonometric function calls and one square root call.

\paragraph*{\bf Iterative method:}
A simple but accurate approximate solution is
\begin{align}
\lambda_{3}={} &\Delta m^2_{31} +\frac1{2}\Delta m^2_{ee}\left(x-1+\sqrt{(1-x)^2+4 x s^2_{13} } ~\right)\,,
\notag \\
\text{with} ~x&\equiv \frac{A_{mat}}{\Delta m^2_{ee}} ~ \text{and} ~ \Delta m^2_{ee} \equiv \Delta m^2_{31}-s^2_{12}\,\Delta m^2_{21} \,,
\label{eq:DMP+}
\end{align}
from \cite{Minakata:2015gra,Denton:2016wmg}.
$\Delta m^2_{ee}$ is the $\Delta m^2$ measured in $\nu_e$ disappearance experiments as defined in \cite{Nunokawa:2005nx}, see also \cite{Parke:2016joa}.
This approximation for $\lambda_{3}$ is an excellent one for both mass orderings 
with a fractional difference to the exact eigenvalue of better than $10^{-4}$ for all neutrino energies, and is exact in vacuum.
Also it contains just one square root and no trigonometric functions.

The precision of eq.~\ref{eq:DMP+} can be further improved to arbitrary precision by using one or more NR iterations\footnote{Other
improvement methods could be used here, such as second order NR, however we use 1st order NR because of its simplicity and rapid rate of convergence.}:
\begin{equation}
\lambda_{3} \rightarrow \lambda_{3} - \frac{\chi(\lambda_{3}) }{ \chi^\prime(\lambda_{3}) }\,,
\label{eq:NR}
\end{equation}
where $\chi$ is the characteristic equation (eq.~\ref{eq:CharX}) and $\chi^\prime= 3\lambda^2 -2A\lambda +B$ is the derivative with respect to $\lambda$ \cite{Denton:2024pzc}.

Once $\lambda_3$ has been determined with sufficient precision from either approach, $\lambda_2$ and $\lambda_1$ can be obtained in a straightforward manner using 
\begin{align}
\Delta \lambda_{21} &= \sqrt{( A-\lambda_{3} )^2-4C/\lambda_{3}}\, , \notag \\
\lambda_{2} &=\frac1{2}( A-\lambda_{3}+ \Delta \lambda_{21})\,, \notag\\
\lambda_{1} &=\frac1{2}( A-\lambda_{3}- \Delta \lambda_{21}) \,,
\label{eq:lambda21}
\end{align}
giving us all three eigenvalues.
This requires only one more square root.
It is possible to determine $\lambda_1$ and $\lambda_2$ using a different pair of $A$, $B$, and $C$ with comparable results; the exact solution to $\lambda_3$ would yield identical values for the other eigenvalues for any pair of $A$, $B$, and $C$.
The choice of $A$ and $C$ tends to minimize the error in $\lambda_1$ and $\lambda_2$.
For both NO and IO, $\lambda_3$ is never equal to 0 (or even parametrically small), so dividing by $\lambda_3$ is never an issue.

\subsubsection{Calculation of Eigenvectors}
For the oscillation amplitudes, what is needed from the eigenvectors are the combinations, $\widetilde V_{\alpha i}\widetilde V_{\beta i}$ (which are all real) for $i=2,3$.
We will use the following identities \cite{Denton:2019ovn, Denton:2019pka,Abdullahi:2022fkh}
\begin{align}
    \widetilde V_{e2}^2 &= \frac{\lambda^2_2 -S_{ee} \lambda_2 + T_{ee}}{\Delta \lambda_{23} \Delta \lambda_{21}} \,, \notag \\
    \widetilde V_{e2}\widetilde V_{\mu 2} &= \frac{-S_{e\mu} \lambda_2 + T_{e\mu}}{\Delta \lambda_{23} \Delta \lambda_{21}} \,, \notag\\
    \widetilde V_{e2}\widetilde V_{\tau 2} &= \frac{-S_{e\tau} \lambda_2 + T_{e\tau}}{\Delta \lambda_{23} \Delta \lambda_{21}}\, .
    \label{eq:Va2}
    \end{align}
    Choosing the positive square root for $\widetilde V_{e2}$, these equations give $(\widetilde V_{e2}, \widetilde V_{\mu2}, \widetilde V_{\tau 2})$, the eigenvector for $\lambda_2$.
The square root can be avoided, since only $\widetilde V_{\alpha 2}\widetilde V_{\beta 2}$'s are actually needed, by noting that
\begin{align}
    \widetilde V_{\mu2}^2 &=(\widetilde V_{e2}\widetilde V_{\mu 2})^2/ \widetilde V_{e2}^2 \,,\notag\\
    \widetilde V_{\tau2}^2 &=(\widetilde V_{e2}\widetilde V_{\tau 2})^2/\widetilde V_{e2}^2 =1-\widetilde V_{e2}^2-\widetilde V_{\mu 2}^2 \,,\notag  \\
    \widetilde V_{\mu2}\widetilde V_{\tau 2} &= (\widetilde V_{e2} \widetilde V_{\mu 2}) (\widetilde V_{e2} \widetilde V_{\tau 2})/ \widetilde V_{e2}^2 \,.
    \label{eq:Va2_2}
\end{align}
The $\widetilde V_{\alpha 3}$'s can be similarly calculated using $\lambda_3$ instead of $\lambda_2$ in the numerator of eq.~\ref{eq:Va2} and replacing the denominator with $\Delta \lambda_{31}\Delta \lambda_{32}$.  
The coefficients $S_{e \alpha}$ and $T_{e \alpha}$ are the first rows of the matrices $S \equiv(2E)( \text{Tr}[\widetilde{H}]I -\widetilde{H})$   and $T \equiv \text{Adj}[(2E)\widetilde{H}]$, see appendix \ref{app:Adj}, and are simply given by
\begin{align}
    S_{ee} &=\Delta m^2_{31} c^2_{13} +\Delta m^2_{21}(1-c^2_{13}s^2_{12})\,, \notag\\
    S_{e\mu}&= -\Delta m^2_{21} c_{13}s_{12} c_{12}\,, \notag\\
    S_{e\tau} & = -(\Delta m^2_{31} -s^2_{12} \Delta m^2_{21})s_{13}c_{13}\,,
    \label{eq:Ss}
\end{align}
and 
\begin{align}
    T_{ee} &=\Delta m^2_{31} \Delta m^2_{21} c^2_{13}c^2_{12}\,, \notag \\
    T_{e\mu}&= -\Delta m^2_{31} \Delta m^2_{21}  c_{13} s_{12} c_{12}\,, \notag \\
    T_{e\tau} & = -\Delta m^2_{31} \Delta m^2_{21}  s_{13}c_{13} c^2_{12}\,.
    \label{eq:Ts}
\end{align}
Note, none of these coefficients depend on the matter potential, $A_{mat}$, so are the same for all layers of the Earth given a fixed set of neutrino parameters, $(\Delta m^2_{31}, \Delta m^2_{21}, s^2_{13}, s^2_{12})$. In the 9  combinations of the oscillation parameters needed, $A$, $B$, $C$, $S_{e\alpha}$, $T_{e\alpha}$, some combinations of parameters appear more than once, such as $s_{13} c_{13}$ and $c_{13}s_{12} c_{12}$, therefore an effort has been made to avoid duplicate computations.

Now we have all the ingredients to calculate the amplitude for neutrinos propagating through one Earth layer of constant density given by
   \begin{align}
    \Re[{\widetilde{\cal A}}_{\alpha \beta} ] = \delta_{\alpha \beta} &+\widetilde V_{\alpha 2}\widetilde V_{\beta 2}(\cos[ \Delta \lambda_{21}L/(2E)] -1) \notag \\
    &+\widetilde V_{\alpha 3}\widetilde V_{\beta 3}(\cos[ \Delta \lambda_{31}L/(2E)]-1) \notag \\
    \Im[\widetilde {\cal A}_{\alpha \beta} ]= &-\widetilde V_{\alpha 2}\widetilde V_{\beta 2}\sin[ \Delta \lambda_{21}L/(2E)] \notag \\
    &-\widetilde V_{\alpha 3}\widetilde V_{\beta 3}\sin[\Delta \lambda_{31}L/(2E)] \,
    \label{eq:amp-layer2} 
\end{align}
where we have explicitly given the real and imaginary parts of $\widetilde {\cal A}$. For each layer this oscillation amplitude is a complex symmetric, $\widetilde {\cal A}_{\alpha \beta}=\widetilde {\cal A}_{\beta \alpha}$, and unitary matrix, $\widetilde {\cal A}\widetilde {\cal A}^\dagger=\widetilde {\cal A}^\dagger\widetilde {\cal A}=\mathbbm1$.

\subsection{Amplitudes to Probabilities}
\label{sec:amplitudes}
Given the inner amplitude in the tilde basis (without the $\theta_{23}$ and $\delta$ rotation) for one layer, we next combine the amplitudes through each layer in the Earth.
We also consider that the trajectories may start in the atmosphere and that the detector may be located inside the Earth.

\subsubsection{Combining the Layer Amplitudes}

\begin{figure}
\centering
\includegraphics[width=0.40\textwidth]{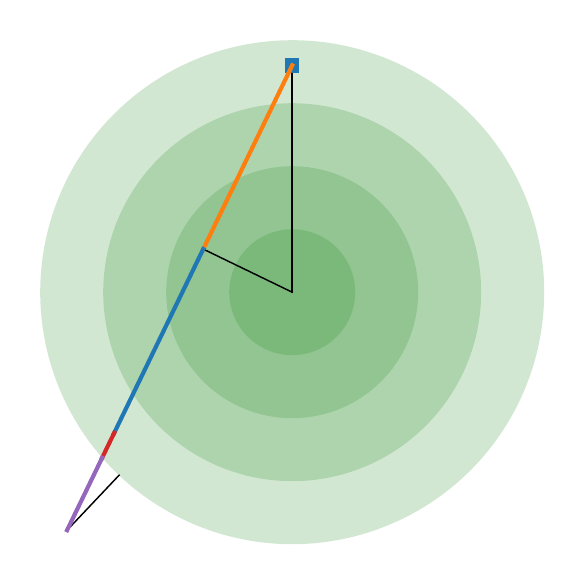}
\caption{A sample trajectory through an Earth model showing the four segments.
First, the purple segment, $A$, is from the production point to the surface.
Second, the red segment, $S$, is from the surface to the detector depth on the far side.
Third, the blue segment, $I$, is from the detector depth on the far side to the deepest point of the trajectory.
Fourth, the orange segment, $O$, is from the deepest point to the detector, the blue rectangle.}
\label{fig:Traj_Sch}
\end{figure}

With the ability to calculate the oscillation amplitude for each layer, construction of the full amplitude from the production, high in the atmosphere, through the Earth to the detector is straight forward. We divide the neutrino trajectory into four segments, see also Fig.~\ref{fig:Traj_Sch}:
\paragraph*{\bf Segment $A$:} From production in the atmosphere to the surface. The amplitude for this segment is called $\widetilde{\cal A}_A$, where $A$ stands for air.
\paragraph*{\bf Segment $S$:} From the surface on the far side to the detector depth on the same side of the Earth. For a trajectory with positive zenith angle this and the $A$ segment will will be the only segments. The amplitude for this segment is called $\widetilde{\cal A}_S$, where $S$ stands for surface.
\paragraph*{\bf Segment $I$:} From the detector depth in the Earth to the point of the trajectory which is closest to the center of the Earth. The amplitude for this segment is called $\widetilde{\cal A}_{I}$, where $I$ stands for in.
\paragraph*{\bf Segment $O$:} From the point closest  to the center  of the Earth of the trajectory to the detector. The amplitude for this segment is called $\widetilde{\cal A}_{O}$, where $O$ stands for out.
    
Each of these segments is then subdivided into multiple layers with different densities and different path lengths in each layer depending on the zenith angle.
This construction is designed in this way such that the oscillation amplitudes in the $I$ and the $O$ segments are the transpose of one another since the density-length profiles are identical except in reverse order which can save up to a factor of two in computational effort.
This construction also factorizes out the propagation in the atmosphere which can be trivially calculated and adjusted for different production heights.

The three amplitudes in the Earth are just products of many one layer amplitudes:
\begin{align}
\widetilde{\cal A}_S &= (\widetilde{\cal A}_{s_m} \cdots \widetilde{\cal A}_{s_1}) \,, \notag \\
\widetilde{\cal A}_I &= (\widetilde{\cal A}_{n} \cdots \widetilde{\cal A}_{1}) \,, \\
\widetilde{\cal A}_O &= (\widetilde{\cal A}_{1} \cdots \widetilde{\cal A}_{n}) = \widetilde{\cal A}_I^{\,T} \,, \notag
\end{align}
where $s_1, \dots s_m$ are labels for the Earth layers between the Earth's surface and the detector depth on the production side, whereas $(1, \dots n)$ label the layers between the detector depth and the layer where the chord is closest to the center of the Earth. 
$\widetilde{\cal A}_{A}$ is the oscillation amplitude between the neutrino production point and the Earth's surface which is easily calculated in vacuum.
As it is on the outside of the matrix it can be changed without changing any of the rest of the calculation.

The fact that $\tilde{\cal A}_O =\widetilde{\cal A}^T_I $ depends on two features; one is that the density profile for 
$\widetilde{\cal A}_I$ is the same as for $\tilde{\cal A}_O$ with inverse order and the other is that each layer amplitude is a symmetric matrix, so that taking the transpose of $\widetilde{\cal A}_I $ gives the correct $\widetilde{\cal A}_O$ amplitude.

The full amplitude matrix for a given trajectory, $\widetilde{\cal A}_{traj}$ is given by
\begin{align}
    \widetilde{\cal A}_{traj} &= \widetilde{\cal A}_O \widetilde{\cal A}_{I} \widetilde{\cal A}_{S}\widetilde{\cal A}_A= \widetilde{\cal A}_I^{\,T} \, \widetilde{\cal A}_{I} \, \widetilde{\cal A}_{S}\widetilde{\cal A}_A\,.
\end{align}
There are a total of $m+n+1$ multiplications of $3\times3$ complex matrices, down from $m+2n+1$ if the identity $\widetilde A_O=\widetilde A_I^T$ was not used.
When the detector is within a few kilometers of the Earth's surface, $m$ is expected to be a small integer and $n$ a much larger integer, especially for trajectories that cross through much of the Earth, leading to nearly a factor of two reduction in computational cost.
If we are using an $N$-layer Earth model, then for any trajectory $m+n \leq N$, with the maximum number of such matrix multiplications occurring for a core crossing trajectory.

At this point $\theta_{23}$ and $\delta$ have not appeared in the calculation of $\widetilde{\cal A}_{traj}$, which is the most computationally expensive part of computing the entire probability.
Thus we can now calculate the oscillation probability with many values of $\sin^2 \theta_{23}$ and $\delta$ very inexpensively.
This is especially useful as these variables are the oscillation parameters which are known the least at the current time, and both play an important role in atmospheric oscillations.

\subsubsection{Transforming back to the flavor basis}
Now we transform our calculation from the tilde basis back to the standard flavor basis.
Note that this step does not depend in any way on the number of layers or how the Earth model is handled.

To include $\theta_{23}$ and $\delta$ and transform back to the flavor basis, we perform the following rotation on the amplitude along the full trajectory through the Earth and the atmosphere $\widetilde {\cal A}_{traj}$:
\begin{align}
{\cal A}_f = R_{23}D_3\widetilde{\cal A}_{traj}D_3^*R_{23}^T\,,
\label{eq:Atilde to Af}
\end{align}
as in eq.~\ref{eq:tildeH to Hf} where, again, $D_3^*(\delta)=D_3(-\delta)$.
Then the oscillation probability matrix, $P$, is given by
\begin{align}
P_{\alpha \beta}= |\, {{\cal A}}_{f}[\alpha, \beta] \, |^2 \,.
\label{eq:Af to P}
\end{align}
It is straightforward to confirm that the sum of each row adds to 1 and the sum of each column adds to 1, due to the unitarity of ${{\cal A}}_{f}$.

Due to this unitary nature of the P matrix, only 4 elements need to be directly calculated from $\tilde{\cal A}_{traj}$, saving on complex computation.
We have determined the simplest independent four; 
\begin{align}
    P_{ee} &= |\widetilde{{\cal A}}_{traj}[1,1] |^2
    \label{eq:Pee}
\end{align}
as the $\delta$ and $\theta_{23}$ rotations do not affect electron neutrino disappearance. 
Two of the appearance channels are simply given by
\begin{align}
    P_{e\mu} &= |\, c_{23} \widetilde{{\cal A}}_{traj}[1,2] +s_{23}\, e^{ -i\delta} \widetilde{{\cal A}}_{traj}[1,3] \,|^2   \label{eq:Pem} \\
    P_{\mu e} &= |\, c_{23} \widetilde{{\cal A}}_{traj}[2,1] + s_{23} \, e^{ +i\delta} \widetilde{{\cal A}}_{traj}[3,1]\, |^2   \label{eq:Pme} \,.
\end{align}
Note that CPV appears here, in the interference terms, due to the $e^{ +i\delta}$ in the [3,i] and $e^{ -i\delta}$ in the [i,3] elements of $D_3\widetilde{\cal A}_{traj}D_3^*$, for i=1 or 2.  The muon neutrino disappearance channel is slightly more complicated, given by
\begin{align}
    P_{\mu\mu} &=|\, c^2_{23} \widetilde{{\cal A}}_{traj}[2,2] + s^2_{23} \widetilde{{\cal A}}_{traj}[3,3] \label{eq:Pmm} \\ 
    & \quad \quad +s_{23} c_{23} ( e^{ -i\delta} \,\widetilde{{\cal A}}_{traj}[2,3]
    +e^{ +i\delta}\,\widetilde{{\cal A}}_{traj}[3,2]) \,|^2\,. \notag   
\end{align}
Eqs.~\ref{eq:Pee}-\ref{eq:Pmm} are identical to those found in refs.~\cite{Yokomakura:2002av, Blennow:2013rca, Ge:2013zua}. Since $\widetilde{{\cal A}}_{traj}$ is unitary, all oscillation probabilities can be calculated using only 4 independent elements of $\widetilde{{\cal A}}_{traj}$. Here, our choices are determined so as to minimize complex arithmetic.

With these four elements of the $P$ matrix,  the other five elements are simply obtained via:
\begin{align}
    P_{e\tau} &= 1-P_{ee}-P_{e\mu} \,,
    \quad P_{\mu\tau}
    = 1-P_{\mu e}-P_{\mu\mu} \,,  \notag \\
    P_{\tau e} &= 1-P_{ee}-P_{\mu e} \,,
    \quad P_{\tau \mu}
    = 1-P_{e\mu}-P_{\mu\mu} \,, \\
    P_{\tau \tau} &= P_{ee}+P_{\mu \mu} + P_{e\mu}+P_{ \mu e} -1 \,, \notag
\end{align}
completing the calculation.
A basic cross-check is that all elements of $P$ must satisfy, $ 0\leq  P_{\alpha \beta} \leq 1$.
When  $\widetilde{{\cal A}}_{traj}$ is unitary, this set of constraints is guaranteed to be satisfied. 

\section{Astrophysical Neutrinos from the Sun or a Supernova}
\label{sec:astro}
The same advances to efficiently calculate atmospheric neutrinos through the Earth also apply naturally to neutrinos from the Sun or a supernova as they pass through the Earth.

\subsection{Solar Neutrinos}
Solar neutrino oscillations can be classified into two categories: daytime and nighttime.
Daytime solar neutrinos are produced as $\nu_e$'s in the dense environment of the Sun immediately in the matter/propagation basis and remain in the same state as they adiabatically propagate out of the Sun \cite{Mikheyev:1985zog}.
They then propagate to the Earth as mass states and decohere and interact in the detector as flavor states.
This is described by the daytime solar neutrino probability
\begin{equation}
P_{e\beta}^\odot=\sum_i|U^\odot_{ei}|^2|U_{\beta i}|^2\,,
\end{equation}
where $U^\odot$ is the unitary matrix that diagonalizes the Hamiltonian at the density in the Sun at the production point, see eq.~\ref{eq:HF}.
This calculation is relatively straightforward and not extremely computationally expensive.
The challenging part is for nighttime neutrinos where the matter effect in the Earth contributes at an important level.

We note that the amplitude for the probability through the Earth from the mass basis, $\mathcal A_{i\beta}^\oplus$ where $P_{i\beta}^\oplus=|\mathcal A_{i\beta}^\oplus|^2$, is related to the one used for atmospheric neutrinos by one additional PMNS matrix.

Nighttime solar neutrinos contain an important regeneration effect.
It is known, based on oscillation probability parameters and some preliminary data, that the solar $\nu_e$ disappearance probability should be slightly higher, on average, at night than during the day.
While the probability oscillates quite quickly, directly observing those oscillations requires energy resolution and statistics far beyond what is achievable.
Nonetheless, the averaged effect modifies the probability at the few \% level and is potentially detectable.
This nighttime probability starts with the same structure as for daytime solar neutrinos, except that instead of immediately projecting from the mass state to the flavor state at the detector, the mass state must propagate through the Earth to the detector.
The nighttime probability is then
\begin{equation}
P_{e\beta}^\odot=\sum_i|U^\odot_{e i}|^2P_{i\beta}^\oplus\,,
\end{equation}
which depends on both energy and zenith angle and where the $\odot$ and $\oplus$ make the solar and terrestrial contexts clear.
In addition, in the limit where the distance traveled in the Earth goes to zero, $P_{i\beta}^\oplus\to|U_{\beta i}|^2$ and we recover the daytime expression.
The computation of $P_{i\beta}^\oplus$ through the Earth for solar neutrinos with \texttt{NuFast-Earth} leverages the same powerful advantages provided for atmospheric neutrinos.

\subsection{Galactic Supernova Neutrinos}
The neutrino signal coming from a galactic supernova will contain a wealth of information about supernova dynamics and neutrino oscillations.
Depending on the time of day and location in the sky, the signal may well travel through the Earth before detection which will modify the signal somewhat \cite{Lagage:1987xu,Arafune:1987cj,Notzold:1987vc,Minakata:1987fj} and the calculation of that probability will need a high degree of precision due to the importance of such a dataset.
This code can also be used for that by using the solar part but setting the production density much higher to quantify the MSW effect in the supernova.
In addition, unlike for solar neutrinos which are essentially only electron neutrinos, all six flavors (including both neutrinos and antineutrinos) are populated in the source requiring the computation of all oscillation channels.

\section{Results}
\label{sec:results}
Here we present the results of running the code in various standard configurations.
We use the nu-fit v6 numbers, for the normal ordering without SK-atmospheric \cite{Esteban:2024eli} unless otherwise noted; see table \ref{tab:osc params}.
For the inverted ordering, we simply switch the sign on $\Delta m^2_{31}$.
We use a detector depth of 2 km and set the production height to 10 km, except for solar neutrinos.
For solar neutrinos we use a production density of $\rho Y_e=100 \times (\frac23$) g/cc.
We use an eigenvalue precision of 1 NR iteration, the \verb|PREM_NDiscontinuityLayer(2,10,10,5)| Earth model, and the \verb|-O3| compiler flag, unless otherwise specified.

\begin{table}
\centering
\caption{Oscillation parameters \cite{Esteban:2024eli} for the NO used in this text.
For the IO the sign of $\Delta m^2_{31}$ is flipped.}
\label{tab:osc params}
\begin{tabular}{c|c|c|c|c|c}
$s_{12}^2$ & $s_{13}^2$ & $s_{23}^2$ & $\delta$ & $\Delta m^2_{21}$ [eV$^2$] & $\Delta m^2_{31}$ [eV$^2$] \\\hline
0.307 &\, 0.02195 &\, 0.561 & 177$^\circ$ &\,  $7.49\e{-5}$ &\, $+2.534\e{-3}$
\end{tabular}
\end{table}

\begin{figure*}
\centering
\includegraphics[width=0.49\textwidth]{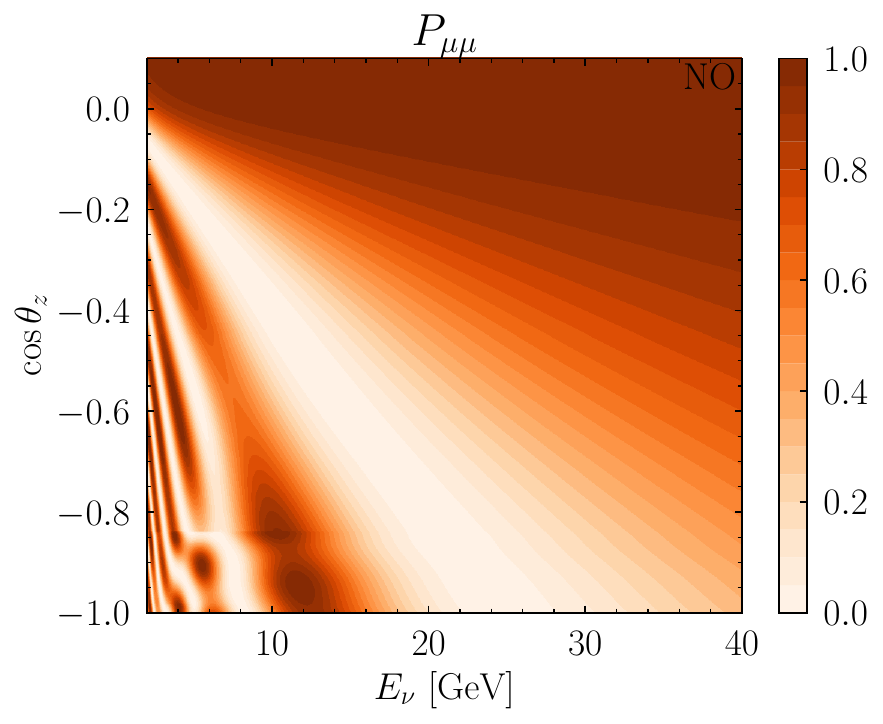}
\includegraphics[width=0.49\textwidth]{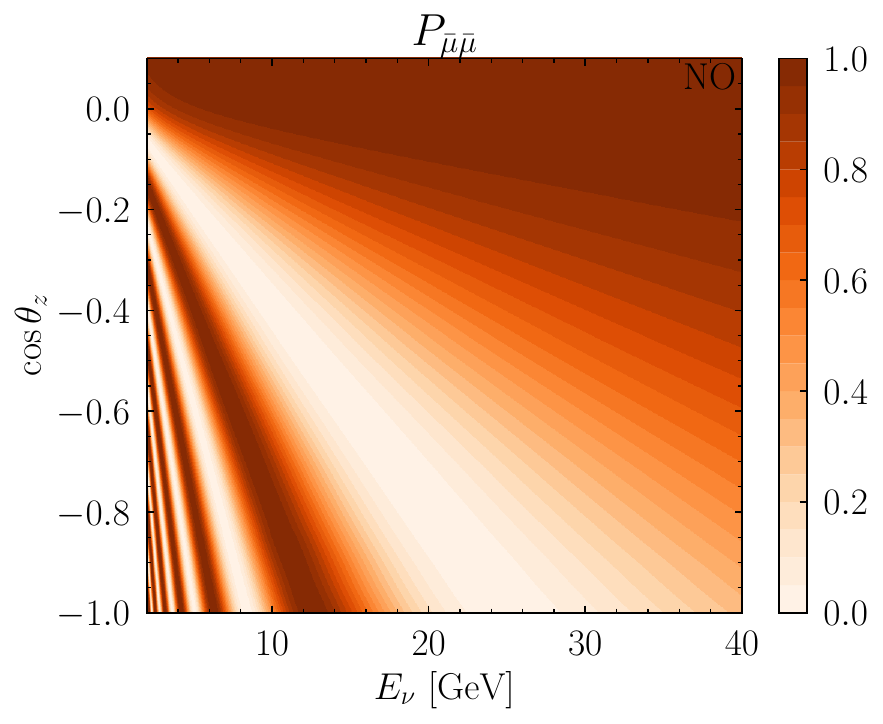}
\includegraphics[width=0.49\textwidth]{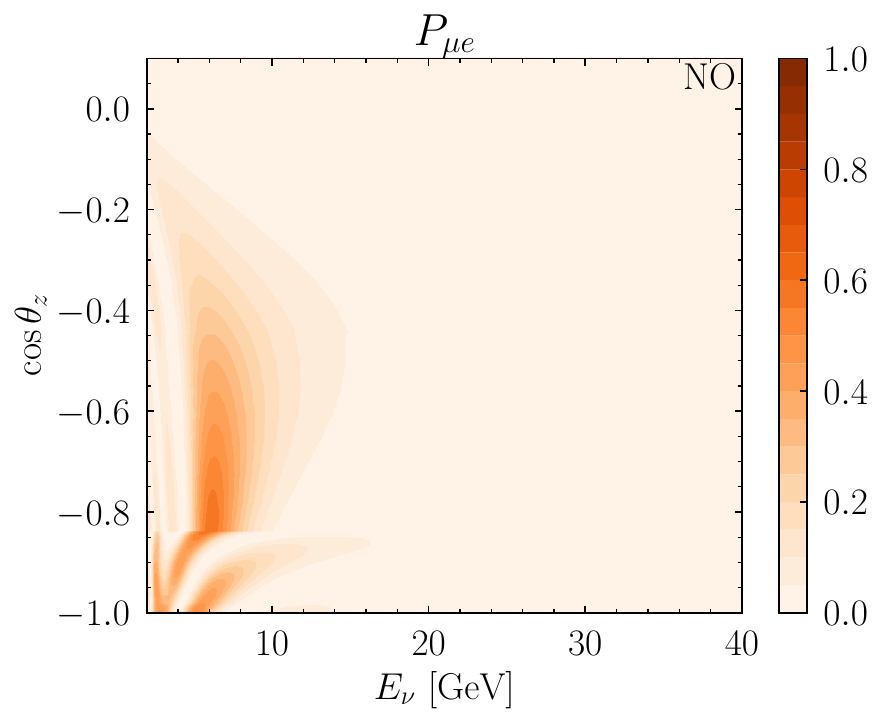}
\includegraphics[width=0.49\textwidth]{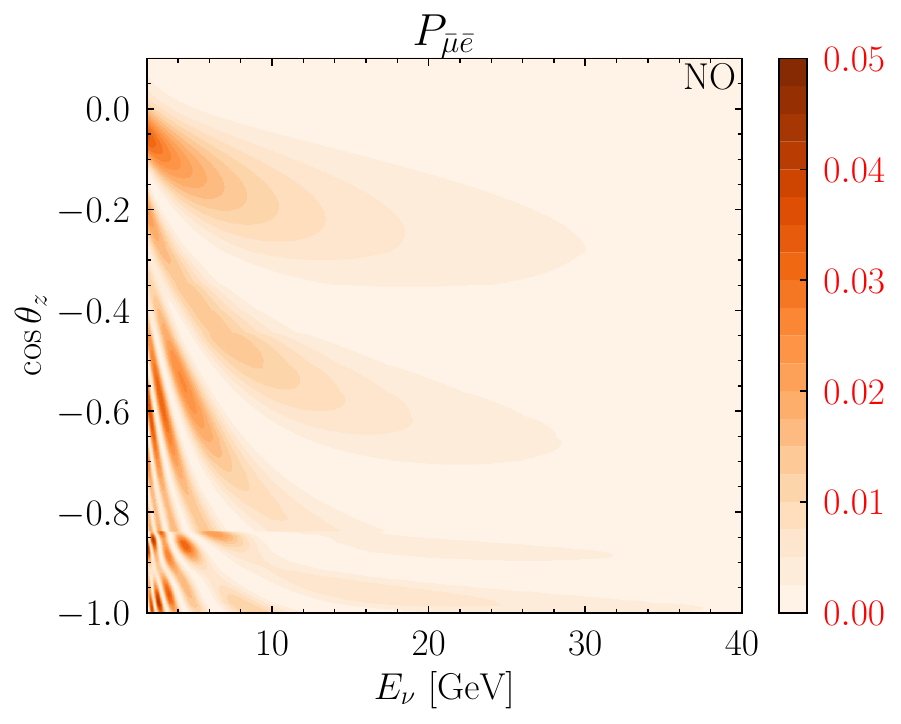}
\caption{The disappearance (\textbf{top}) and appearance (\textbf{bottom}) probabilities across parameters typical for atmospheric parameters in the normal ordering.
The \textbf{left} (\textbf{right}) panels are for neutrinos (antineutrinos).
The color scale for the appearance figure for antineutrinos is smaller to highlight hard to see features. 
The detector depth is 2 km and the production height is 10 km.}
\label{fig:oscillogram NO}
\end{figure*}

\begin{figure*}
\centering
\includegraphics[width=0.49\textwidth]{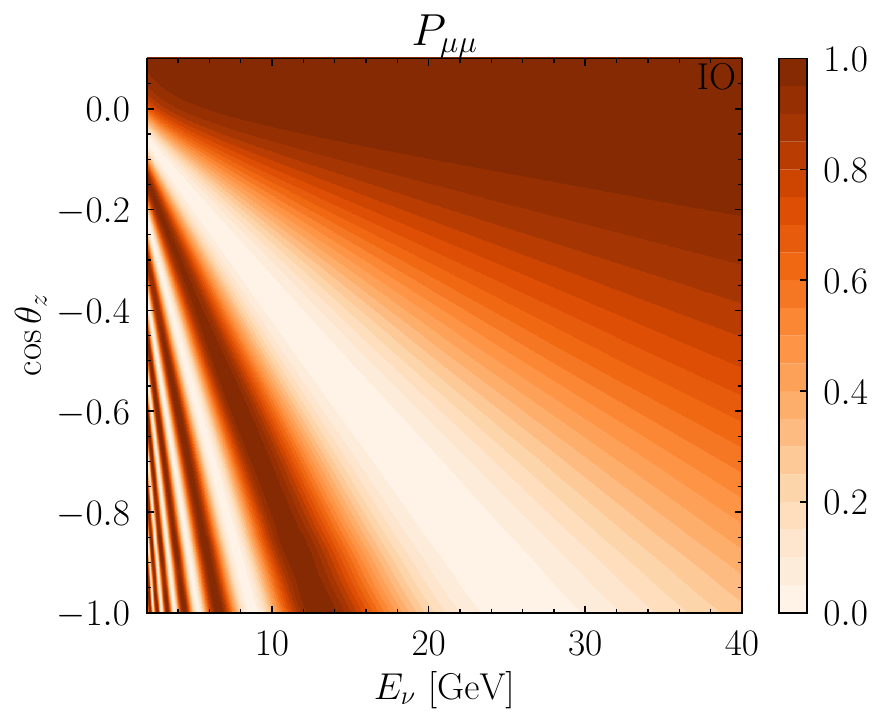}
\includegraphics[width=0.49\textwidth]{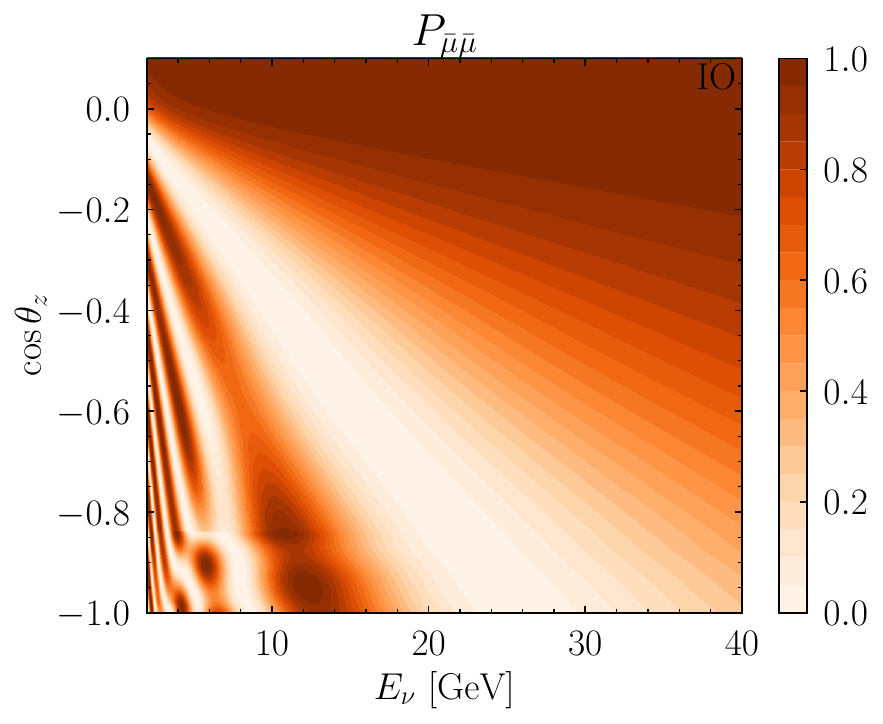}
\includegraphics[width=0.49\textwidth]{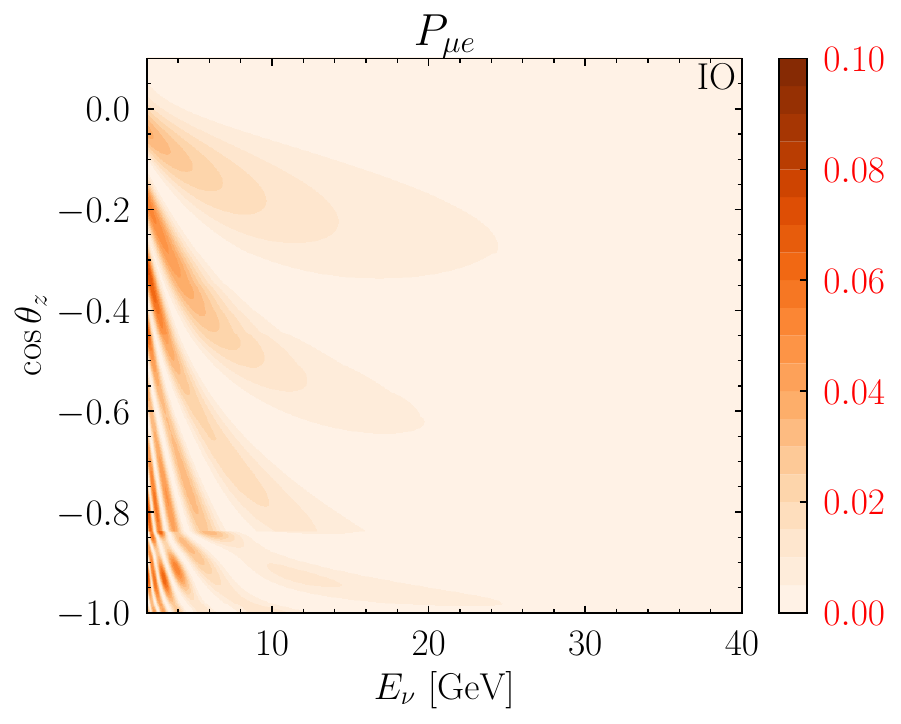}
\includegraphics[width=0.49\textwidth]{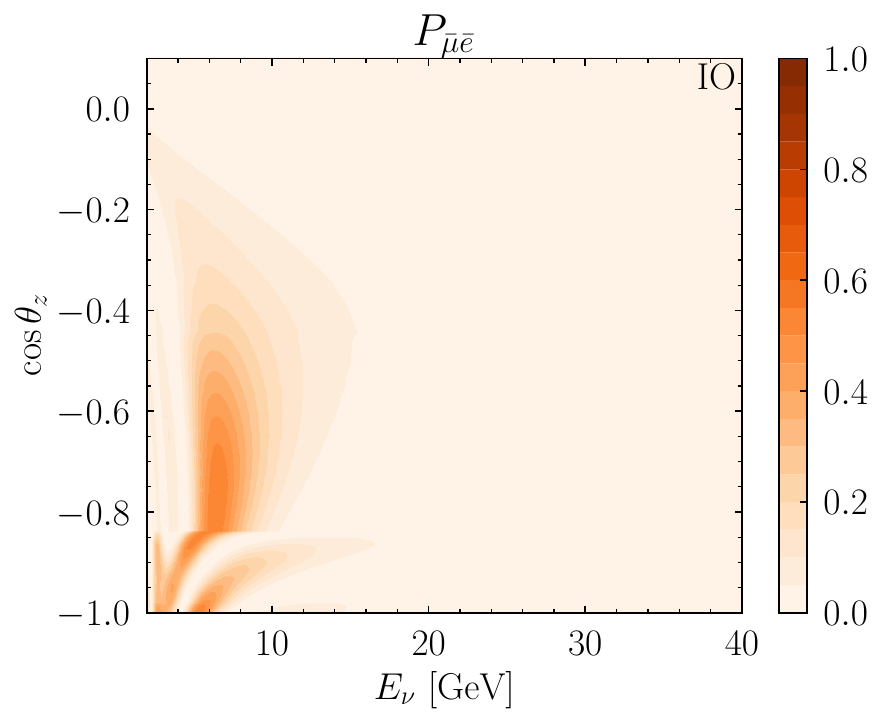}
\caption{The same as Fig.~\ref{fig:oscillogram NO} but in the inverted ordering.
The color scale for the appearance figure for neutrinos is smaller to highlight hard to see features.}
\label{fig:oscillogram IO}
\end{figure*}

Figures \ref{fig:oscillogram NO}-\ref{fig:oscillogram IO} show the oscillograms for atmospheric neutrino oscillation probabilities as a function of energy and zenith angle in both disappearance and appearance channels, for neutrinos and antineutrinos, and for the normal and inverted mass orderings.
These highlight the well known parametric resonance through the Earth's core that depends on the mass ordering \cite{Petcov:1998su,Akhmedov:1998xq}.
The impact of nonzero detector depth can be seen in that for low energy neutrinos, $\lesssim$ few GeV, we see a deviation from $P_{\mu\mu}=1$ at $\cos\theta_z>0$.

We also investigated the role of detector depth by computing the difference in the oscillation probability for a detector on the surface and a detector 2 km below the surface.
We summarized the results in table \ref{tab:detector depth} which shows that, for atmospheric neutrinos, the variation in the probability is up to a few percent in some channels.
Oscillograms of all 8 channels are included in the appendix for completeness, see Figs.~\ref{fig:detector depth NO}-\ref{fig:detector depth IO}.
These show that, in most cases, the effect is for horizontal to slightly down-going trajectories at low energies, which is a fairly small fraction of phase space and does not contribute much to the oscillation information which tends to come from trajectories through the mantle and/or the core.
Nonetheless, for precise calculations, including the nonzero detector depth is essential.

\begin{table}
\centering
\caption{Maximum shift in probability over $E\in[2,40]$ GeV and $\cos\theta_z\in[-1,0.1]$ between a detector on the surface and a detector at a depth of 2 km.}
\begin{tabular}{|c|c|c|c|c|}
\hline
\multirow{2}{*}{$\max|\Delta P|$} & \multicolumn{2}{c|}{$\nu_\mu\to\nu_\mu$} & \multicolumn{2}{c|}{$\nu_\mu\to\nu_e$}\\\cline{2-5}
& $\nu$ & $\bar\nu$ & $\nu$ & $\bar\nu$ \\\hline
NO & 5.1\% & 5.1\% & 1.5\% & 0.2\%\\\hline
IO & 5.4\% & 5.4\% & 0.3\% & 1.5\%\\\hline
\end{tabular}
\label{tab:detector depth}
\end{table}

\begin{figure}
\centering
\includegraphics[width=\columnwidth]{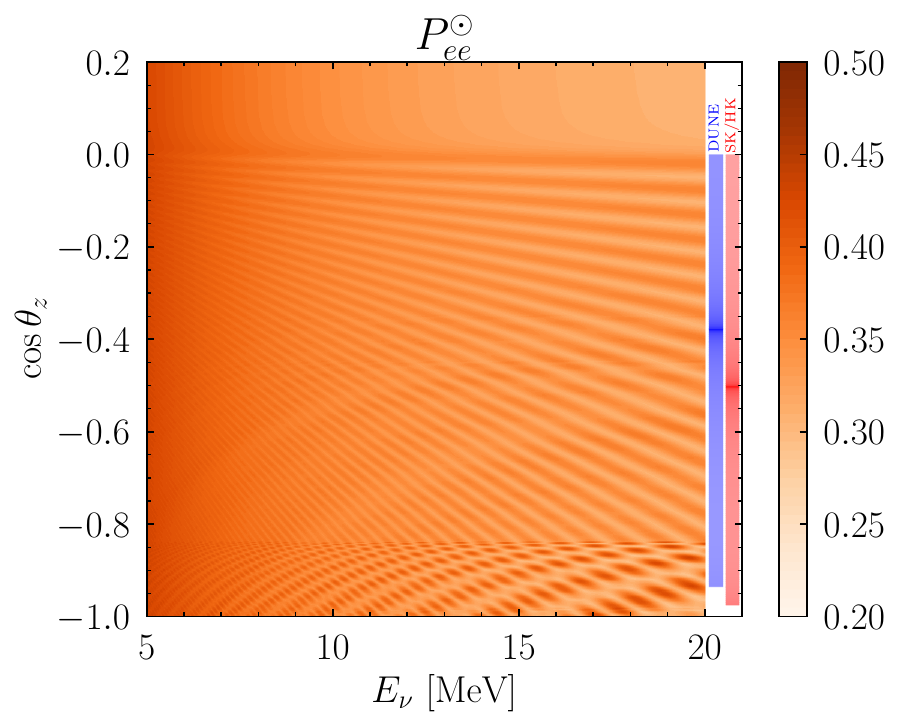}
\caption{The $\nu_e$ disappearance probability from the center of the Sun to a detector in the Earth as a function of energy and zenith angle.
The bands on the right show the annual average exposure for nighttime neutrinos at the latitudes for DUNE (blue) and SK/HK (red).
The detector depth is at 2 km which leads to some matter effects for daytime ($\cos\theta_z>0$) neutrinos as well.}
\label{fig:solar oscillogram}
\end{figure}

Next, we compute the $\nu_e\to\nu_e$ disappearance probability for solar neutrinos shown in Fig.~\ref{fig:solar oscillogram}.
We have focused on the higher energy, $E>5$ MeV, part of the spectrum where the impact of the matter effect starts to become most relevant.
The red and blue bands show the average annual nighttime exposure weighted in terms of $\cos\theta_z$ modified from \cite{Lisi:1997yc}, see also Fig.~\ref{fig:solar weight validation} in the appendix.
It is easy to see the significant impact the matter effect has on the core crossing trajectories.

\begin{figure}
\centering
\includegraphics[width=\columnwidth]{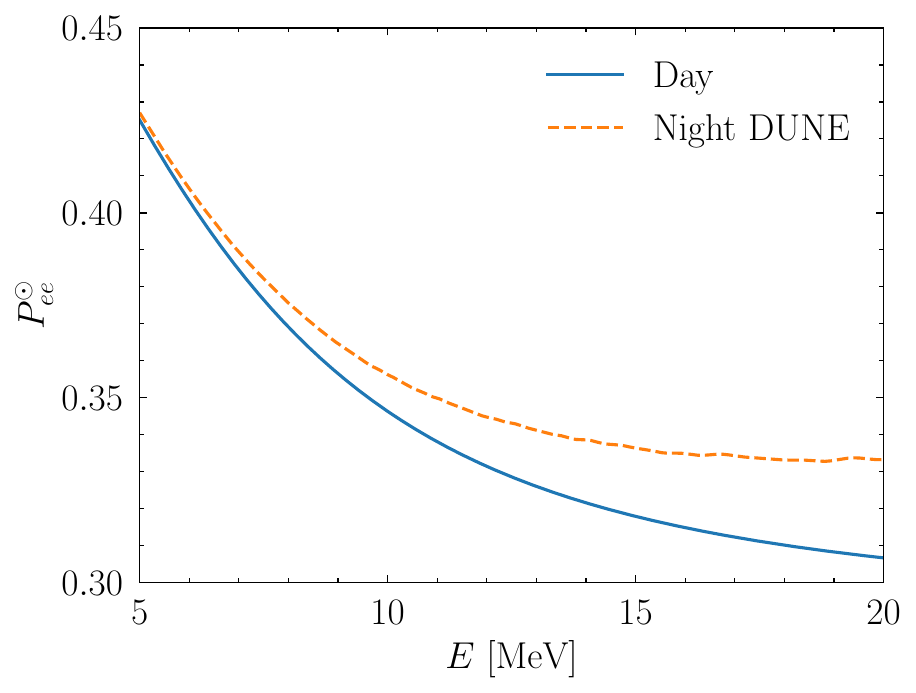}
\caption{The daytime solar $\nu_e$ disappearance probability (blue) and the nighttime solar disappearance probability (orange dashed) averaged over a year of exposure at DUNE's latitude.}
\label{fig:solar day night validation}
\end{figure}

In Fig.~\ref{fig:solar day night validation} we show the annual averaged nighttime probability for DUNE's latitude along with the daytime probability with no terrestrial matter effect, clearly showing the effect of regeneration, see e.g.~\cite{Akhmedov:2004rq}.
For further validation examples, see appendix \ref{sec:validation}.

\section{Code Usage}
\label{sec:code}
For details of using the code, see the user's guide on github \cite{nufast-earth-github}.
We do provide one minimal working example for atmospheric neutrinos here:\\\vspace*{0.05in}

\begin{lstlisting}
// Initialize the engine
Probability_Engine probability_engine;

// Set the oscillation parameters to nu-fit 6 best fit values:
probability_engine.Set_Oscillation_Parameters(0.307, 0.02195, 0.561, 177 * M_PI / 180, 7.49e-5, 2.534e-3, true); //(s12sq, s13sq, s23sq, delta, Dmsq21, Dmsq31, neutrino_mode)

// Set energy and zenith angle arrays
std::vector<double> energies = {1, 2, 3, 4, 5}; // GeV
std::vector<double> coszs = {-1, -0.5, 0, 1}; // core-crossing to horizontal to down-going
probability_engine.Set_Spectra(energies, coszs);

// Create Earth model instance
PREM_NDiscontinuityLayer earth_density(2, 10, 10, 5); // 2 layers in the inner core, 10 layers in the outer core, 10 layers in the inner mantle, and 5 layers in the outer mantle
// Set Earth details
probability_engine.Set_Earth(2, &earth_density); // detector depth in km
// Set production height (optional)
probability_enginge.Set_Production_Height(10); // km, recalculations after changing this are fast

// Do the calculations
std::vector<std::vector<Matrix3r>> probabilities = probabilitie_engine.Get_Probabilities();
\end{lstlisting}

After the calculations are done, the user will likely be interested in changing some parameters such as oscillation parameters, the production height in the atmosphere, or details about the Earth density model.
One of the key advantages of the algorithm described here and the implementation in the code is that certain parameters are considered ``fast'' for which minimal recomputations are required, while others require recomputing the entire trajectory through the Earth.
The following parameters are ``fast'':
\begin{gather*}
\bullet ~ \text{mixing angle} \, \theta_{23} ~ \bullet ~ \text{CP phase} \, \delta\\ 
\bullet \, \text{Production height}~\bullet\,\text{Solar density}
\end{gather*}
Any set of nested loops over a set of parameters that contains one or more of these four parameters should have these parameters on the innermost loops.

\section{Speed, Precision, and Recommendations}
\label{sec:speed}
In this section we benchmark our implementation of the algorithm and provide recommendations for usage.
For any particular usage case the details may differ and the details may also differ depending on the architecture the code is run on.
Everything can be parallelized in a variety of ways, but we leave these details to the user.

\subsection{Speed}
First, we compare the speed of each step of the computations.
We do this on a Thinkpad laptop with an i7 chip using \texttt{gcc} version 13.3.
We consider three different compiler flags: \texttt{-O0} which contains no optimizations and is an example of the slowest simplest version, \texttt{-O3} which is very slightly aggressive but generally considered to be safe, and \texttt{-Ofast} which is quite aggressive and may bring some additional speed over \texttt{-O3}, but may clash with some third-party codes.
Unless specified otherwise, our default run configuration is \texttt{-O3} and with eigenvalue precision of 1.

\begin{figure}
\centering
\includegraphics[width=\columnwidth]{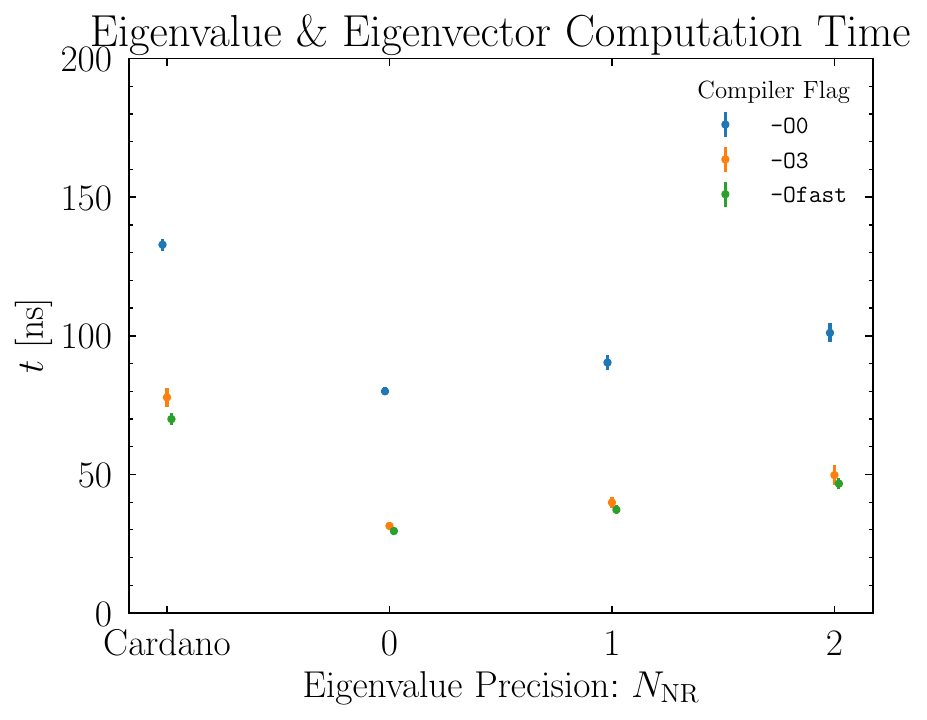}
\caption{The mean computation time to compute the necessary eigenvectors and eigenvalues with the exact expressions (Cardano) and using the approximate expression, eq.~\ref{eq:DMP+}, with 0, 1 and 2 NR iterations ($N_{\rm NR}$).
The colors indicate different compiler flags.
Note that at $N_{\rm NR}=2$ the accuracy is equivalent to Cardano using double precision arithmetic.}
\label{fig:speed eigen}
\end{figure}

We benchmark the time it takes to compute one set of three eigenvalues and corresponding eigenvectors under the different compiler flags and for different levels of precision.
We compute the time it takes to compute the eigenvalues and eigenvectors for a standard set of parameters 10,000 times to get an estimate of the time it takes to compute the eigenvalues once.
We then repeat this 10,000 times to get an uncertainty due to variations in the computer's state.
Finally, we repeat the procedure for different compiler flags.
Our results are shown in Fig.~\ref{fig:speed eigen} which shows that even at an eigenvalue precision of 2, which essentially reaches the double precision limit (see Fig.~\ref{fig:precision} below), the approximation technique is still definitively faster than the exact expression from Cardano \cite{cardano}.
Our recommended precision is 1 (although 0 is also sufficient for all practical uses as well) as that is significantly more precise than will be needed for next generation atmospheric and solar measurements and provides a factor of $\sim2$ speedup in the computing the eigenvalues and eigenvectors over the exact expression.
We also recommend a compiler flag of \verb|-O3| as it is significantly faster than \verb|-O0| and \verb|-Ofast| offers no real advantage with some additional compatibility issues.

\begin{figure}
\centering
\includegraphics[width=\columnwidth]{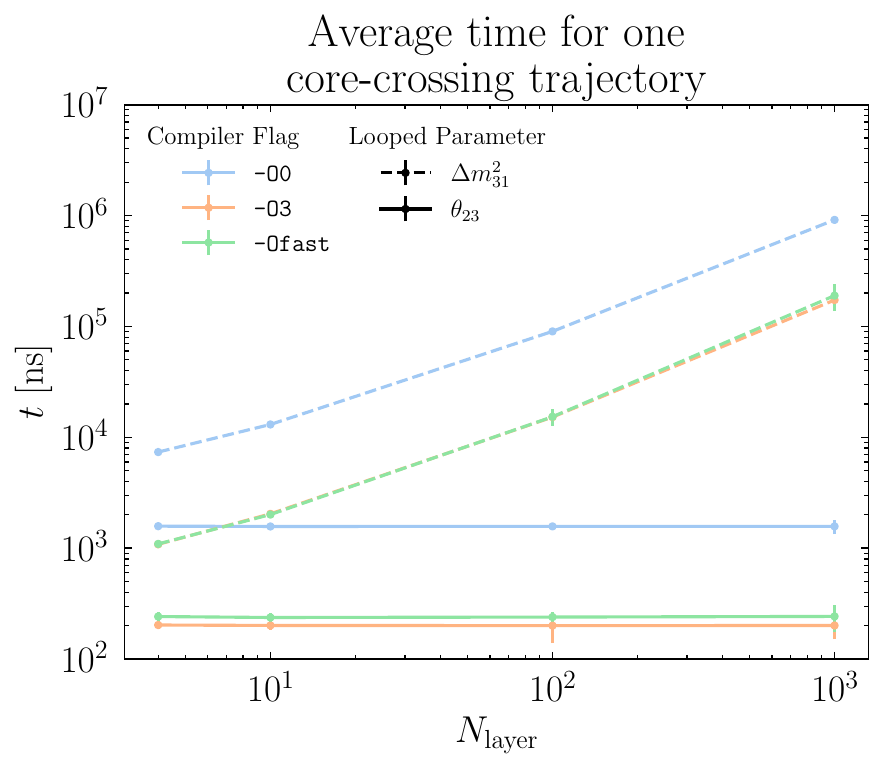}
\caption{The mean computation time to compute the probabilities through the Earth's core for a varying number of layers and with different compiler flags.
The solid and dashed lines represent loops where either $\theta_{23}$ or $\Delta m^2_{31}$ is changed each time, respectively.}
\label{fig:atmospheric speed}
\end{figure}

We combine everything into the total time required to compute one core crossing trajectory.
We use the \verb|PREM_NUniformLayer| with varying numbers of layers for simplicity.
We compute the probabilities 10,000 times and take the mean time.
We then repeat this 100 times to get an estimate of the mean and the uncertainty.
The numerical results for different compiler flags are shown in Fig.~\ref{fig:atmospheric speed}.
The results are shown where either $\theta_{23}$ is changed for each of the 10,000 computations or $\Delta m^2_{31}$ is changed each time.
We see that changing $\theta_{23}$ is faster at any Earth model and the cost to change it does not increase with the complexity of the Earth model, as it does for $\Delta m^2_{31}$.
Similar fast results are achieved where changing $\delta$ or the production height, while changing the other oscillation parameters, $\theta_{12}$, $\theta_{13}$, and $\Delta m^2_{21}$ require recomputing all the amplitudes through the Earth.
This numerically shows that the so-called ``fast'' parameters, $\theta_{23}$, $\delta$, and the production height can be varied for essentially no significant computational cost regardless of the complexity of the Earth model.

\begin{figure}
\centering
\includegraphics[width=\columnwidth]{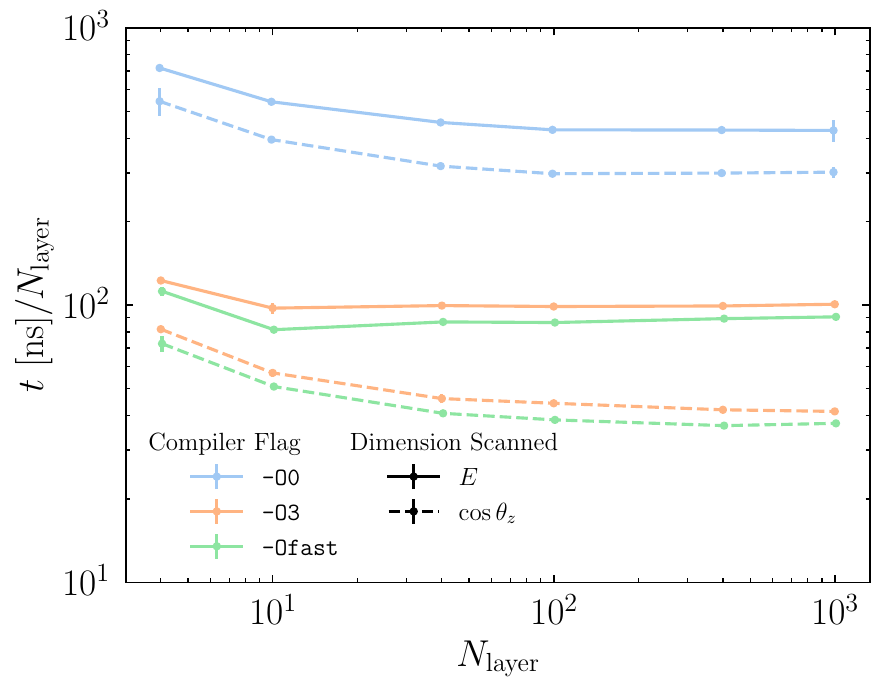}
\caption{The mean computation time for one core-crossing trajectory divided by the number of layers for different numbers of layers and for different compiler flags.
The solid and dashed lines represent cases where the algorithm is used at 100 grid points in energies and zenith angles, respectively. Clearly, zenith angle scans are approximately twice as fast as energy scans using compiler flags \texttt{-03} and \texttt{-0fast}.}
\label{fig:E vs cosz speed}
\end{figure}

The time to compute probabilities across many zenith angles and across many energies differs between these two parameters.
Specifically, increasing the number of zenith angles grid points is cheaper than increasing the number of energy grid points with the difference being about a factor of 2, see Fig.~\ref{fig:E vs cosz speed}.
The difference is more modest for Earth models with $\lesssim30$ layers before it asymptotes to this factor of $\sim2$.
We have confirmed that this trend holds quite generally regardless of the number of zenith angle or energy grid points, as well as the compiler options.
The reason for this benefit is that the eigenvalues and eigenvectors need to be calculated for each layer in the Earth and for each density.
Thus varying the zenith angle many times does not lead to an increase in the number of eigenvalue and eigenvector computations while varying the energy many times does.
Both cases still require computing the amplitudes, multiplying them together, and then performing the final transformation from $\widetilde{\mathcal A}_{traj}$ to the probabilities, see Eqs.~\ref{eq:Atilde to Af}-\ref{eq:Af to P}.
What this result shows is that latter part of the computation takes a comparable amount of time as computing all the eigenvalues and eigenvectors.

We also show the time it takes to compute the geometry for one trajectory through the Earth in table \ref{tab:trajectory speed}, a step that needs to be performed once for a given Earth model and zenith angle selection.
We compute a single core-crossing trajectory for a variety of the presented Earth models.
We note that the \verb|PREM_Full| and \verb|PREM_Four| models perform an integration along the trajectory so that the number of steps is minimized while maintaining accuracy for a variety of zenith angles.
This step in the calculation is only recalculated if the Earth model or the zenith angles are changed.
The mean time is first estimated by repeating the calculation 1000 times and then repeating that 1000 times provides for an error estimate on the mean.

\begin{table}
\centering
\caption{The average time (ns) to compute the geometry for a single trajectory.
Note that \texttt{PREM\_Full} and \texttt{PREM\_Four} contain an integral across the 10 and 4 large regions, respectively.}
\label{tab:trajectory speed}
\begin{tabular}{c|c|c}
Model & Model parameters & Time (ns)\\\hline
\verb|PREM_NDiscontinuityLayer| & (2, 10, 10, 5) & $396\pm30$\\
\verb|PREM_NDiscontinuityLayer| & (25, 25, 25, 25) & $3499\pm39$\\
\verb|PREM_NUniformLayer| & (4) & $41\pm0.5$\\
\verb|PREM_NUniformLayer| & (10) & $93\pm2$\\
\verb|PREM_NUniformLayer| & (100) & $3501\pm123$\\
\verb|PREM_prob3| & - & $43\pm6$\\
\verb|PREM_Full| & - & $6090\pm206$\\
\verb|PREM_Four| & - & $5802\pm68$
\end{tabular}
\end{table}

We now compare the speed of our code to other popular atmospheric neutrino oscillation codes in the literature.
We benchmarked the speed of \texttt{nuSQuIDS} \cite{Arguelles:2021twb} across a range of energies and zenith angles.
We noted that the code computed some trajectories ($\cos\theta_z$ closer to zero) much faster than others.
We found that the average time to compute one probability was $1.6\e{6}$ ns.
We also note that while this calculation did not include new physics or neutrino interactions in the Earth, the \texttt{nuSQuIDS} code in general is designed to handle these scenarios.
We also benchmarked \texttt{OscProb} \cite{joao_coelho_2025_17425187} and found that for a core-crossing trajectory with 44 layers, the time to compute one probability is $\sim40\,000$ ns, regardless of the oscillation parameter looped over, compared with $\sim6000$ ns when looping over a slow parameter and $\sim200$ ns when looping over a ``fast'' parameter in our \texttt{NuFast-Earth} code.

\subsection{Precision}
\begin{figure}
\centering
\includegraphics[width=\columnwidth]{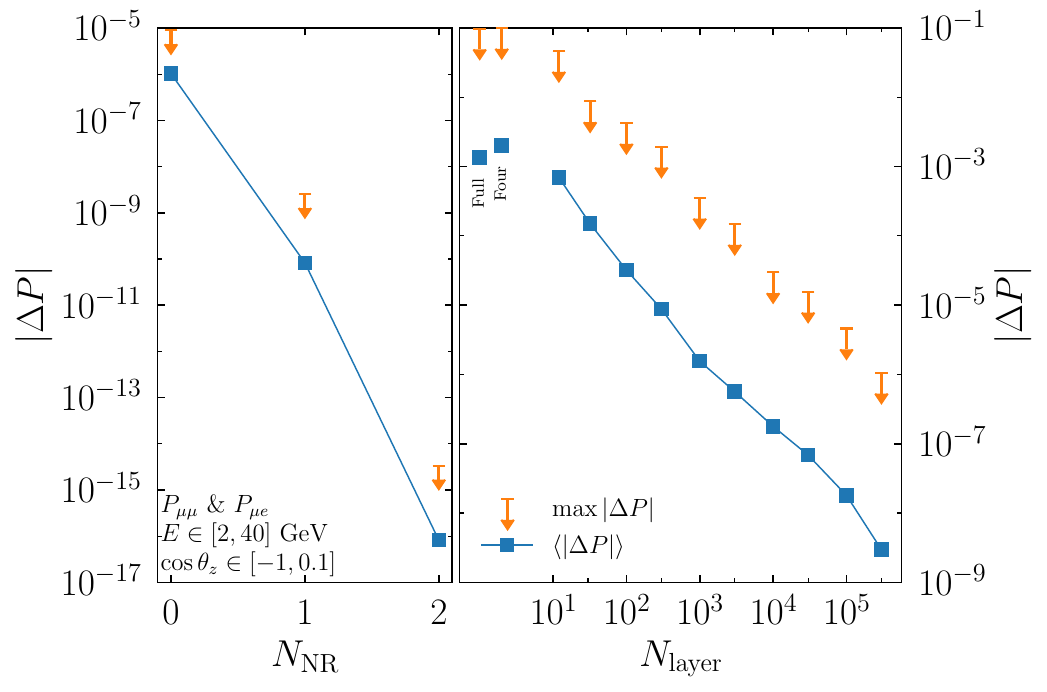}
\caption{The mean error (blue square) and maximum error (orange arrow) in the probabilities across a $100\times100$ grid in energy and zenith angles including both disappearance and appearance.
The two approximations considered are shown: the eigenvalue precision, $N_{\rm NR}$, on the left and the number of layers in the Earth density model, $N_{\rm layer},$ on the right with different vertical scales.
The integration methods \texttt{PREM\_Full} and \texttt{PREM\_Four} are also shown in the right panel.}
\label{fig:precision}
\end{figure}

We consider the precision in terms of the two main approximations used to compute atmospheric oscillation probabilities: the eigenvalues and the Earth density model.
To estimate the error, we compute the oscillation probabilities across a $100\times100$ grid of energies $E\in[2,40]$ GeV and zenith angles $\cos\theta_z\in[-1,0.1]$.
We consider our ``exact'' oscillation probabilities as those with exact eigenvalues and one million total layers, spread out with 250,000 in each of the four main regions via the \verb|PREM_NDiscontinuityLayer| \verb|(2.5e5, 2.5e5, 2.5e5, 2.5e5)| functional call.
We then recalculate the oscillation probabilities across this grid with approximate eigenvalues or fewer layers in the Earth model or with the two integration based Earth models: \verb|PREM_Full| and \verb|PREM_Four|.
Then, to compare the two grids, we compute the maximum and average difference across both of the two main oscillation probabilities: $P(\nu_\mu\to\nu_\mu)$ and $P(\nu_\mu\to\nu_e)$.
The grid scanned is designed to cover the majority of the relevant oscillation physics for atmospheric neutrinos.
Certainly using a finer (coarser) spaced grid would find more (fewer) points with a larger error, but the mean error remains relatively unchanged and the $100\times100$ sized grid is reasonably fine grained for practical applications.
Our results are summarized in Fig.~\ref{fig:precision}.
We find that the mean error as defined above as a function of the total number of layers in the Earth model is well approximated by
\begin{equation}
\langle|\Delta P|\rangle\simeq\frac{0.002}{N_{\rm layer}}\,,
\end{equation}
which provides a useful scaling rule when deciding on the desired precision.

We see that the precision rapidly improves as the number of layers is increased.
At 1000 total layers the maximum error is $<10^{-3}$ and the mean error is $\sim10^{-6}$ which is significantly more precise than will be needed in the coming experiments.
In reality, $\sim32$ total layers with subpercent maximum error and $10^{-4}$ mean error is sufficient for nearly all practical uses.
While this study kept the layers evenly divided among the four regions, further optimization there may be possible.

We also tested the \verb|PREM_Full| and \verb|PREM_Four| Earth density models and find that errors as large as $\sim10\%$ do happen, although the mean error is $\lesssim10^{-3}$ which is acceptable for many cases.

We find that even with zero NR iterations on the eigenvalues, the eigenvalues are already incredibly precise introducing corrections no larger than $10^{-5}$ at the probability level, far below what can be measured.
We also see that NR iterations to the eigenvalues yield vast further improvements in the probabilities all the way down to the double precision limit with just two iterations.

\subsection{Recommendations}
There are a number of choices of how to implement neutrino oscillations through the Earth and we carefully go through our recommendations here.

First we address the choice for the eigenvalues.
While it may be tempting to simply set this to the exact expression (by calling any negative number in the \verb|Set_Eigenvalue_Precision| function), this is not necessary.
We see in Fig.~\ref{fig:precision} that the difference between two NR and the exact expression is indistinguishable and Fig.~\ref{fig:speed eigen} shows that computing the eigenvalues and eigenvectors is over a third faster when using two NR iterations than when using the exact expression.
Thus we never recommend using the exact Cardano expression.
In fact, given the level of precision achieved in the approximations, we find it unlikely that an experiment can even differentiate between zero NR iterations and one NR iteration.
Our official recommendation is to use one NR iteration and that is what the code uses by default, but zero NR iterations should not be distinguishable and is slightly faster.

Second is the more complicated question of the Earth models.
Some have proposed using the integration technique implemented in the \verb|PREM_Full| or the slightly simpler \verb|PREM_Four| Earth density models.
When comparing Earth models with a very small ($\lesssim10$) layers, this approach may be beneficial.
It addresses the problem of neutrinos just skimming into e.g.~the outer core and seeing the average density of that region when the actual density seen is less due to the modest variation within that region.
We find that the advantages of this approach do not outweigh the costs, which are significant.
The initial integration time is slower (see table \ref{tab:trajectory speed}) but this is not likely to be significant.
The main issues are twofold.
First, we find that it introduces a correction compared to the model of using many steps.
Second, it requires computing the eigenvalues and eigenvectors for each energy, each layer, \emph{and each zenith angle}.
So while the number of layers may be smaller, if the number of zenith angles is not small then it may be considerably more expensive than using many layers.
Thus we recommend one of the constant density layered approaches.
We have found that the uniform spacing of layers can lead to some problems near the sharp boundaries so we always recommend \verb|PREM_NDiscontinuityLayer| over the similar (but also simpler) \verb|PREM_NUniformLayer|.
The number of layers needed does depend on the precision desired as shown in Fig.~\ref{fig:precision}.
For many applications \verb|PREM_NDiscontinuityLayer(2,10,10,5)| should be sufficient.
These numbers are chosen as the inner core does not vary too much and not many neutrinos pass through it (2), the outer core varies a lot (10), the inner mantle also has a sizable variation and experiences a large amount of the flux (10), and the outer mantle is a smaller region but with more fine grained features (5).
If a high precision calculation is needed, increasing these layers by some factor (e.g.~10) should be suffice.
Further optimization on the split among the regions may yield additional small improvements in speed and/or precision.

Beyond the recommendations for the approximations, we again reiterate the most important recommendation, regardless of the choices made above.
When looping over parameters, we strongly encourage that $\theta_{23}$, $\delta$, the production height, and the density in the Sun, be in the innermost loops, if they are varied in a given calculation.
This will lead to dramatic speed ups as nearly all of the calculations can be reused.
The speed ups are the most dramatic for the most complicated Earth models as changing those parameters only require recomputing steps at the very beginning and/or the very end of the trajectory, but none of the interior steps and thus do not depend on the number of Earth layers.
Even if this is not possible, e.g.~in a Markov chain approach, some steps in the fast directions will still be fast, as will the integrals over the production height (atmospheric) or production region (solar).

We note that the atmosphere is extremely well approximated as vacuum, an approximation which provides considerable speed up as any integral over production height becomes essentially free.
Locating the detector at the correct detector depth is important as this can shift the probability $>1\%$, especially for horizontal trajectories.

Finally, we comment on energy versus zenith angles.
Since the eigenvalues and eigenvectors must be recomputed for each layer and each energy, a large number of energies requires many computations of the eigenvalues and eigenvectors.
If a comparable overall precision can be achieved via interpolation with a shift to more zenith angles and fewer energies in the array, then this can save computational time as described in Fig.~\ref{fig:E vs cosz speed}.

\section{Conclusions}
\label{sec:conclusions}
Current generation neutrino oscillation experiments already require the computation of neutrino oscillation probabilities many times filling up super computers for days.
Next generation experiments achieving higher statistics will require orders of magnitude more throws necessitating improved algorithms for computing neutrino oscillation probabilities efficiently.
We have previously addressed this problem for long-baseline oscillations \cite{Denton:2024pzc}.
To handle the significantly more complicated problem of neutrinos traversing the Earth with its sharply varying density profile coming from the atmosphere, the Sun, and a supernova, requires a more sophisticated approach.
Using some of the same neutrino theory and linear algebra results along with new components necessary to address the Earth's sharply varying density profile, we efficiently propagate neutrinos through the Earth in a specific basis that does not depend on $\theta_{23}$ or $\delta$ applying these parameters only at the beginning and end of the trajectory.
This and other features allow us to efficiently save and reuse computations as parameters are varied.

We have developed, implemented, and tested a new algorithm called \verb|NuFast-Earth| for computing neutrino oscillation probabilities through varying density profiles such as the Earth.
The code is publicly available \cite{nufast-earth-github} with an associated user's guide.

The algorithm and code include several innovative components:
\begin{itemize}
\item The eigenvalues are calculated using an approximate, but extremely precise technique that is quite fast.
\item The eigenvectors are calculated directly from the eigenvalues using state-of-the-art linear algebra advances to optimize speed.
\item The majority of the calculations through the Earth are done without the $\theta_{23}$ and $\delta$ parts of the PMNS matrix.
This makes nearly all the calculations real and simplifies many steps.
\item The propagation through the Earth is largely symmetric reducing the number of computations by nearly two for many Earth layers.
\item The code intelligently saves often repeated calculations leading to significant speed ups.
For example, nearly all of the calculation is unchanged when $\theta_{23}$ or $\delta$ changes; looping over those two variables is extremely fast.
A similar speed up is also implemented for variations in the production height (atmospheric) or density in the Sun (solar neutrinos).
\item Eigenvalue and eigenvector calculations are also cached for each energy and Earth layer, allowing for rapid computations over zenith angle.
\end{itemize}
All of the physics, linear algebra, and coding optimization features combine to one or more orders of magnitude speed up over existing state-of-the-art options.

\section*{Acknowledgements}
We thank Christoph Ternes for helpful comments in the early stages of this project, the anonymous referee for helpful comments, and Taku Izubuchi for a careful examination of the code.
PBD acknowledges support from the US Department of Energy under Grant Contract DE-SC0012704.
This manuscript has been authored in part by FermiForward Discovery Group, LLC under Contract Number 89243024CSC000002 with the U.S. Department of Energy, Office of Science, Office of High Energy Physics.

\appendix

\section{Determinant and Adjugate of \texorpdfstring{$(\lambda I -H)$}{lambdaI-H}:}
\label{app:Adj}

For a $3\times3$ Hermitian matrix, the determinant and adjugate of $(\lambda I -H)$ are given by 
\begin{align}
    &\text{Det}[\lambda I- H] = \notag \\
    & \quad \lambda^3 - \text{Tr}[H]\lambda^2  
    + \frac1{2}(\text{Tr}^2[H]-\text{Tr}[H^2])\lambda      -\text{Det}[H] \,, \notag\\
    &\text{Adj}[\lambda I- H] = \lambda^2 I - (\text{Tr}[H]I-H)\lambda + \text{Adj}[H] \,.
\end{align}
Here $H$ has been rescaled by $(2E)$, compared to eq.~\ref{eq:H}, for this appendix only so 
\begin{align}
    H & \equiv R_{13}(\theta_{13}) R_{12}(\theta_{12})~ M^2
    ~ R^T_{12}(\theta_{12}) R^T_{13}(\theta_{13}) +W   \notag
    \end{align}
    with
    \begin{align}
    M^2&=\text{Diag}(0, \Delta m^2_{21},\Delta m^2_{31}), \quad W=A_{mat}\text{Diag}(1,0,0)\,,  \notag \\
&R_{13}(\theta_{13}) R_{12}(\theta_{12})=  \left( \begin{array}{ccc}
 c_{13}c_{12}  & c_{13}s_{12}&  s_{13} \\
  -s_{12} &  c_{12} &  0\\
-s_{13}c_{12} & -s_{13}s_{12}& c_{13} 
\end{array}
\right) \,.
\notag
 \end{align}
Writing out $H$ explicitly, which is real and symmetric, we have

\begin{align}
H&= \Delta m^2_{31} \left( \begin{array}{ccc}
 s^2_{13}  & 0&  s_{13}c_{13} \\
  . &  0 &  0\\
. & .& c^2_{13} 
\end{array}
\right)
 \notag \\
&+ \Delta m^2_{21} \left( \begin{array}{ccc}
  c^2_{13}s^2_{12} & c_{13} s_{12} c_{12}& -s_{13} c_{13} s^2_{12} \\
. & c^2_{12} & - s_{13} s_{12} c_{12}\\
. & . & s^2_{13}s^2_{12}
\end{array}
\right) \notag \\
&+ A_{mat} \left( \begin{array}{ccc}
  1& 0 & 0\\
  .& 0& 0\\
  .&.&0 
\end{array}
\right) \,.
\end{align}
For simplicity, since $H$ is symmetric the elements below the diagonal, represented by ``.'', are identical to the corresponding elements above the diagonal. We use the same simplifying notation for the symmetric $S$ and $T$ matrices
given below.

Then
\begin{align}
    A &\equiv \text{Tr}[H] = \Delta m^2_{31}+\Delta m^2_{21}+A_{mat} \,, \notag \\
    B &\equiv \frac1{2}( \, \text{Tr}^2[H]-\text{Tr}[H^2] \,)
     \notag \\[2mm]
&=\Delta m^2_{31}\Delta m^2_{21} +  
A_{mat}( \Delta m^2_{31} c^2_{13}+\Delta m^2_{21}(1-c^2_{13} s^2_{12})) \,,  \notag \\[2mm]
    C &\equiv \text{Det}[H] =  A_{mat}\, \Delta m^2_{31}\Delta m^2_{21} c^2_{13} c^2_{12} 
    \,.
\end{align}
These are the coefficients, $A$, $B$, $C$ of the characteristic equation given in eq.~\ref{eq:ABC}.\\

For the adjugate of $(\lambda I -H)$ we need the  real symmetric $S\equiv \text{Tr}[H]I -H$  and $T\equiv \text{Adj}[H]$ matrices \cite{Abdullahi:2022fkh},
\begin{align}
    S &\equiv \text{Tr}[H]I -H   \notag \\
    &= \Delta m^2_{31} \left( \begin{array}{ccc}
 c^2_{13}  & 0&  -s_{13}c_{13} \\
  . &  1 &  0\\
. & .& s^2_{13} 
\end{array}
\right)
 \notag \\
&+ \Delta m^2_{21} \left( \begin{array}{ccc}
 1- c^2_{13}s^2_{12} ~~& -c_{13} s_{12} c_{12}& s_{13} c_{13} s^2_{12} \\
. & s^2_{12} &  s_{13} s_{12} c_{12}\\
 . &  . & ~~1-s^2_{13}s^2_{12}
\end{array}
\right) \notag \\
&+ A_{mat} \left( \begin{array}{ccc}
  0& 0 & 0\\
  .& 1& 0\\
  .&.&1 
\end{array}
\right) \,.
\end{align}
Note that $\text{Tr}[S]=2\text{Tr}[H]$ as required.
The only row that does not involve $A_{mat}$ is the first row, which gives eq.~\ref{eq:Ss}. This is why the e-row was chosen when calculating the $\widetilde V$'s, eq.~\ref{eq:Va2}. All rows and columns give the same eigenvectors, as can be easily checked, and  are given here for completeness and cross checks. That is,
\begin{align}
\widetilde V_{\mu2}^2 &= \frac{\lambda^2_2 -S_{\mu \mu} \lambda_2 + T_{\mu \mu}}{\Delta \lambda_{23} \Delta \lambda_{21}} \,, \quad
\widetilde V_{\tau2}^2 = \frac{\lambda^2_2 -S_{\tau \tau} \lambda_2 + T_{\tau \tau}}{\Delta \lambda_{23} \Delta \lambda_{21}} \,, \notag \\
& \hspace{2cm}  \widetilde V_{\mu 2}\widetilde V_{\tau 2} = \frac{-S_{\mu \tau} \lambda_2 + T_{\mu \tau}}{\Delta \lambda_{23} \Delta \lambda_{21}} \,, 
\end{align}
can be shown to be identical to eq.~\ref{eq:Va2_2}, using the appropriate elements of the S, T matrices and the fact that $\lambda_2$ satisfies the characteristic equation, eq.~\ref{eq:CharX}.

Again for the real symmetric matrix $T \equiv \text{Adj}[H]$, the only row that does not involve $A_{mat}$ is the first row, which gives eq.~\ref{eq:Ts},
as

\begin{align}
    T &\equiv \text{Adj}[H] \notag \\
    &= \Delta m^2_{31} \Delta m^2_{21} \left( \begin{array}{ccc}
 c^2_{13} c^2_{12}  & -c_{13} s_{12} c_{12}  &  -s_{13}c_{13}c^2_{12} \\
  .  &  s^2_{12}  &  s_{13} s_{12} c_{12}\\
 . & . & s^2_{13} c^2_{12}
\end{array}
\right)
 \notag \\
&+ A_{mat}\, 
\Delta m^2_{21} \left( \begin{array}{ccc}
  0& 0 & 0\\
  .&  s^2_{13} s^2_{12} & s_{13}  s_{12} c_{12} \\
  .&   . &  c^2_{12} 
\end{array}
\right) \notag \\
&+ A_{mat}\, \Delta m^2_{31} \left( \begin{array}{ccc}
 0&0&0\\
. & c^2_{13} & 0 \\
. & .  & 0
\end{array}
\right)\,.
\end{align}
This is easy to derive\footnote{The adjugate of a product of matrices is the product of the adjugates of the individual matrices but in reverse order. Also $\text{Adj}[R^T_{ij}] =R_{ij}$ and the adjugate of a $2\times2$ matrix, $X$, is simply given by $\text{Adj}[X] = \text{Tr}[X]I - X$. } by noting that
\begin{equation}
\text{Adj}[M^2]=\Delta m^2_{31} \Delta m^2_{21}\, \text{Diag}(1,0,0)\,,
\end{equation}
therefore the coefficient of $\Delta m^2_{31} \Delta m^2_{21}$ is just 
$R_{13} R_{12}\,\text{Diag}(1,0,0)\,R^T_{12} R^T_{13}$. The coefficients of $A_{mat}\, \Delta m^2_{21}$
and $A_{mat}\, \Delta m^2_{31} $ are obtained from
\begin{align}
&\text{Adj}\left( \begin{array}{cc}
c^2_{12}&  -s_{13}s_{12}c_{12}\\
. 
& s^2_{13}s^2_{12}
\end{array} \right)
= \left(\begin{array}{cc}
s^2_{13}s^2_{12}&  s_{13}s_{12}c_{12}\\
. 
& c^2_{12}
\end{array} \right), \notag \\[2mm]
&\text{Adj}\left(\begin{array}{cc}
0& 0\\
. & c^2_{13}
\end{array} \right) 
=\left(\begin{array}{cc}
c^2_{13}& 0\\
. & 0
\end{array} \right),
\end{align}
 where the elements below the diagonal (``.'') are identical to the corresponding elements above the diagonal as all these matrices are symmetric. 

As  cross checks $H \cdot \text{Adj}[H] =  \text{Adj}[H]\cdot H = \text{Det}[H] I$ and also $\text{Tr}[T]=B$, since for a $3\times3$ matrix the adjugate can be written as 
\begin{align}
    \text{Adj}[H] &= H^2- \text{Tr}[H]H +\frac1{2}(\text{Tr}^2[H]-\text{Tr}[H^2])I \,, \notag \\
      \text{Tr}[\text{Adj}[H]] &= \frac1{2}(\text{Tr}^2[H]-\text{Tr}[H^2]) \equiv B \,.
\end{align}
The coefficients of $\lambda$ in $ \text{Det}[\lambda I- H]$ and $ \text{Adj}[\lambda I- H]$ for an $n\times n$ matrix, are connected, as given by the 
Le~Verrier-Faddeev algorithm, see appendix C of \cite{Abdullahi:2022fkh} for more details.

\section{Validation}
\label{sec:validation}
Here we provide plots validating that the code behaves as expected.

\begin{figure*}
\centering
\includegraphics[width=0.49\textwidth]{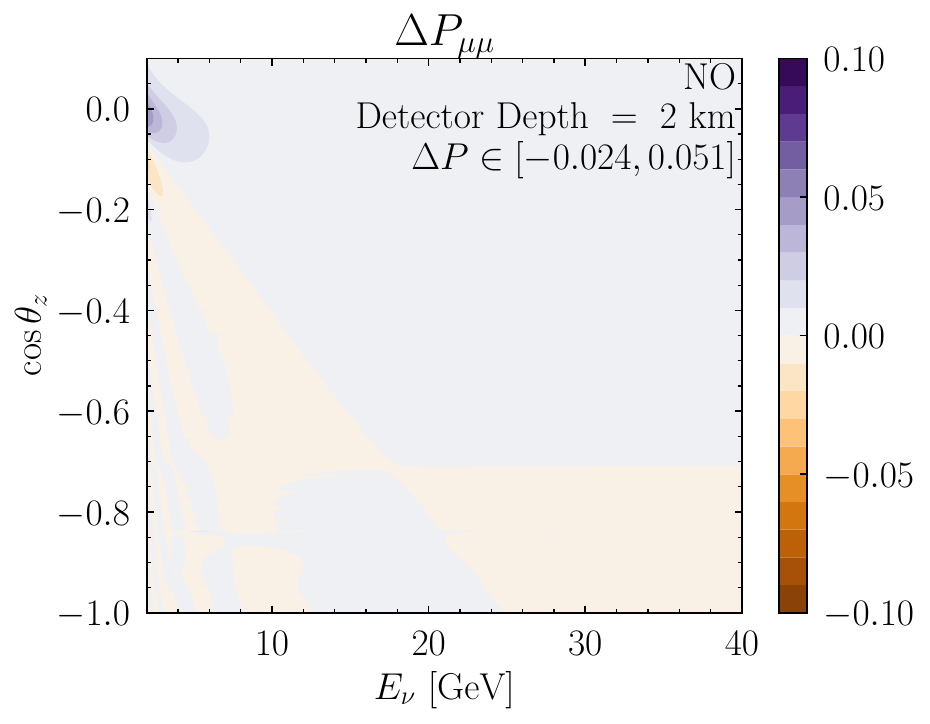}
\includegraphics[width=0.49\textwidth]{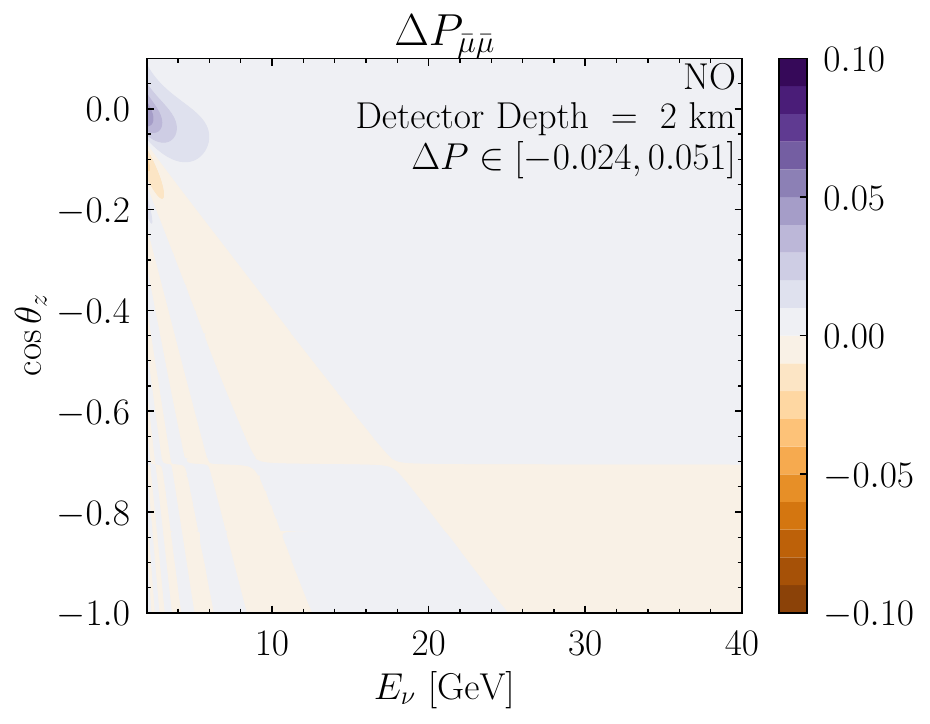}
\includegraphics[width=0.49\textwidth]{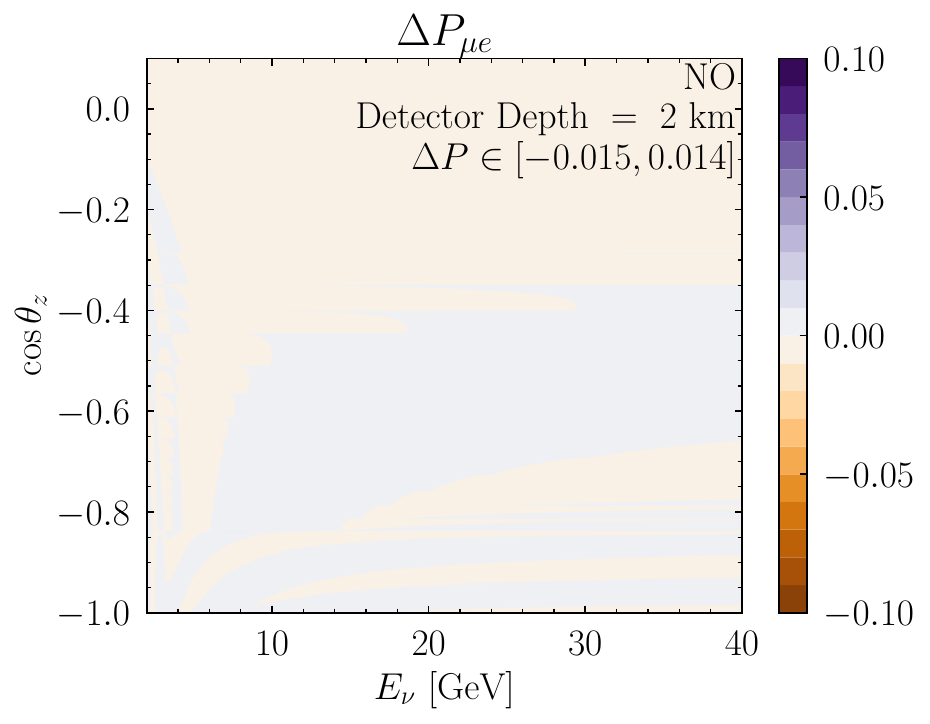}
\includegraphics[width=0.49\textwidth]{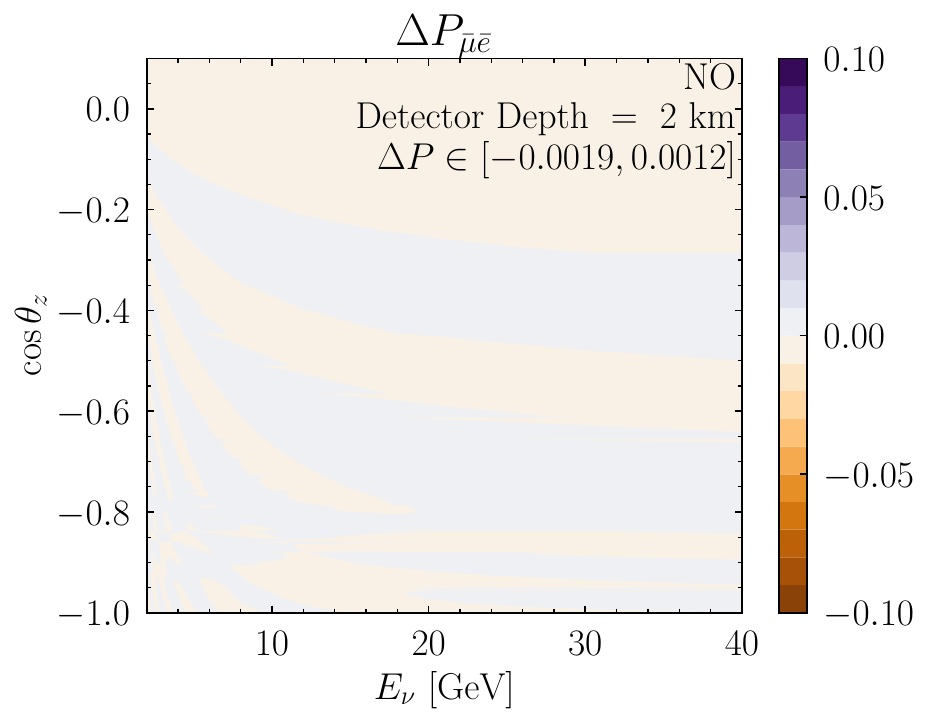}
\caption{The difference in probabilities between a detector on the surface and a detector 2 km below the surface for disappearance (\textbf{top}) and appearance (\textbf{bottom}) probabilities across parameters typical for atmospheric oscillations in the normal ordering.
The inverted ordering is shown in Fig.~\ref{fig:detector depth IO}.
The \textbf{left} (\textbf{right}) panels are for neutrinos (antineutrinos).
The production height is 10 km.
The largest effect is for low energies and near horizontal trajectories.}
\label{fig:detector depth NO}
\end{figure*}

\begin{figure*}
\centering
\includegraphics[width=0.49\textwidth]{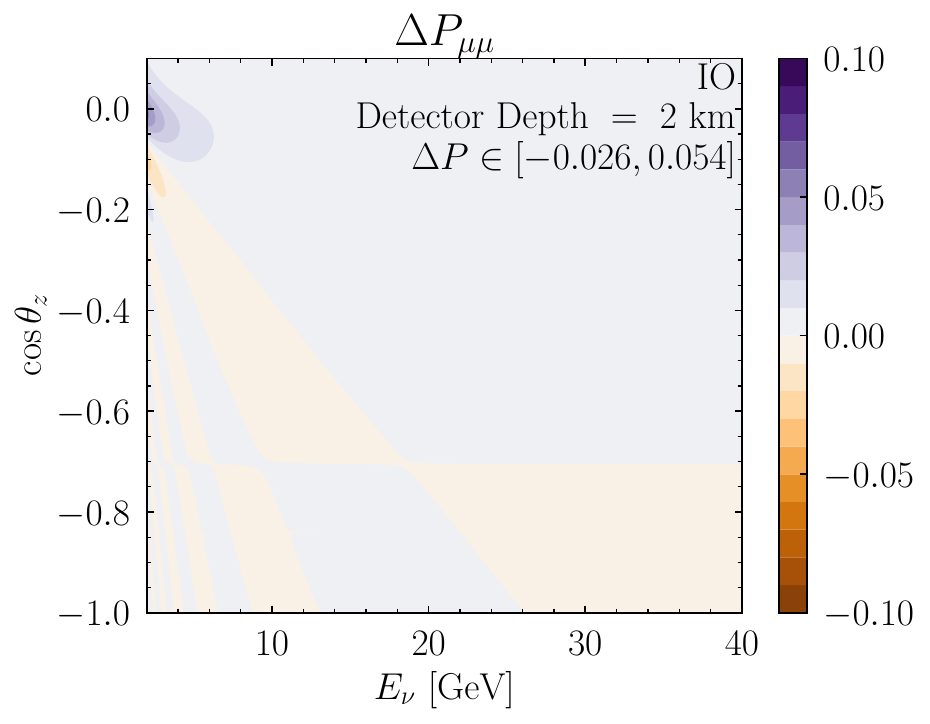}
\includegraphics[width=0.49\textwidth]{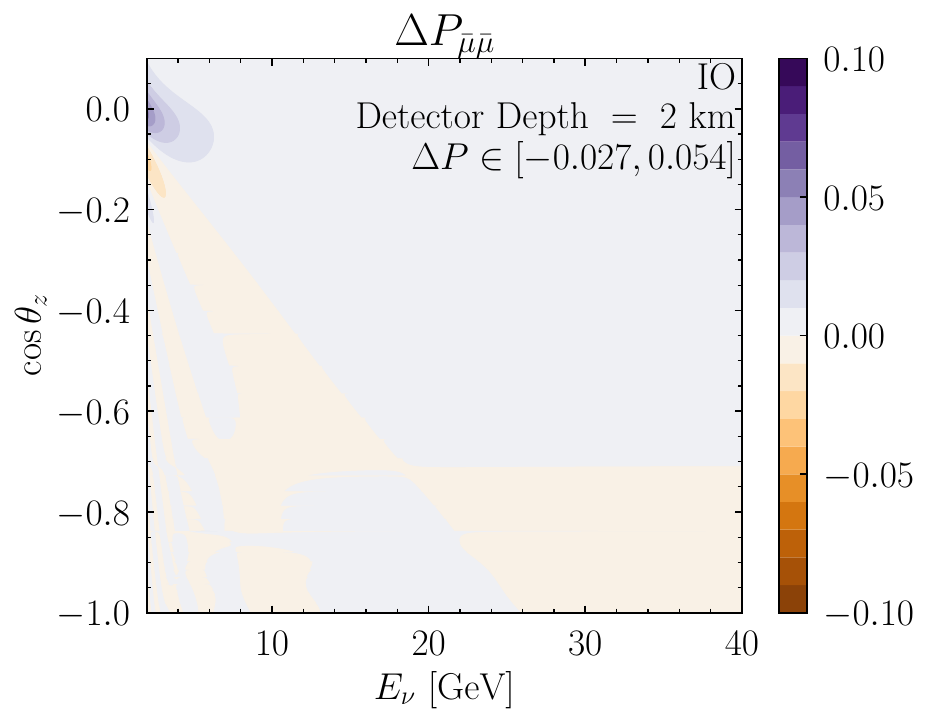}
\includegraphics[width=0.49\textwidth]{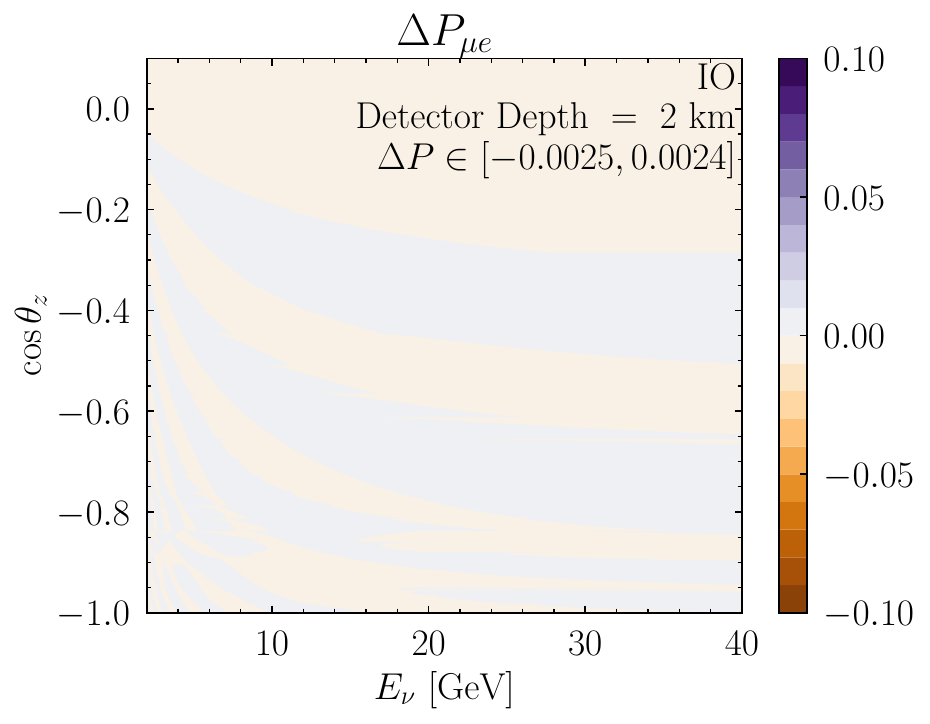}
\includegraphics[width=0.49\textwidth]{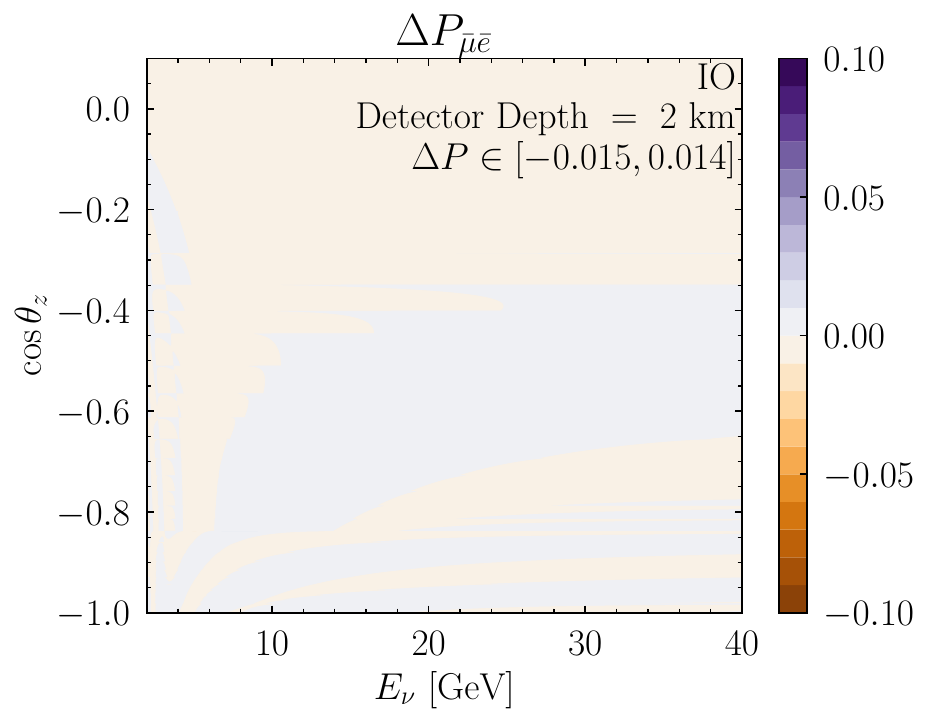}
\caption{The same as Fig.~\ref{fig:detector depth NO} but in the IO.}
\label{fig:detector depth IO}
\end{figure*}

In Figs.~\ref{fig:detector depth NO}-\ref{fig:detector depth IO} we show the difference in probability between a detector on the surface and a detector 2 km deep for all four relevant oscillation channels in both the NO and the IO.
The dominant features are horizontal trajectories at low energies.
See also table \ref{tab:detector depth} for a summary of these figures.

\begin{figure}
\centering
\includegraphics[width=\columnwidth]{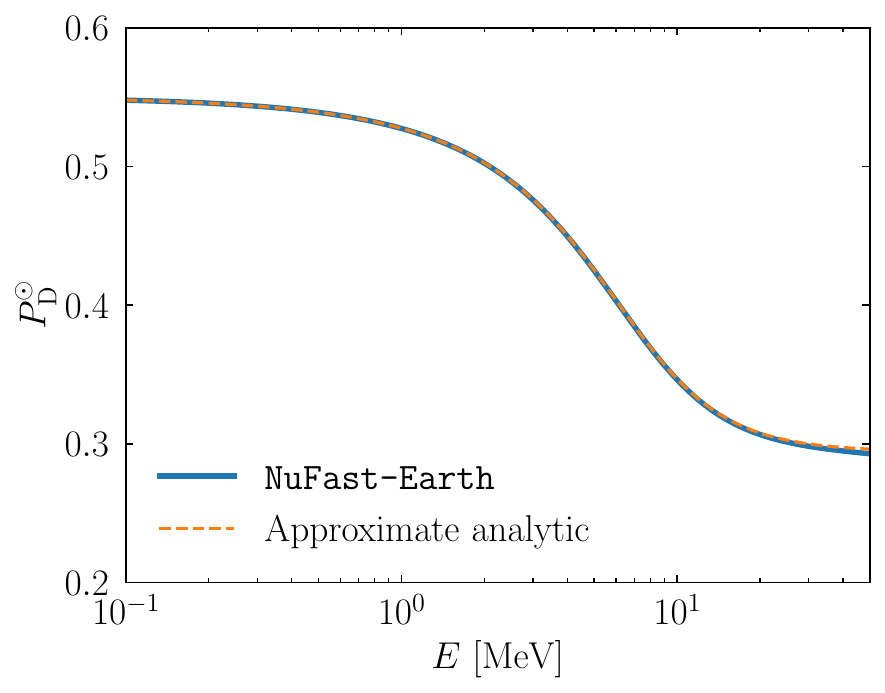}
\caption{A comparison between the numerical result from this \texttt{NuFast-Earth} code for solar neutrinos and the approximate analytic expression from eq.~\ref{eq:PeeD} in the text.
The small difference at high energies is due to the beginnings of the $\Delta m^2_{31}$ effect which is not encompassed in the approximate expression.}
\label{fig:solar day validation}
\end{figure}

In Fig.~\ref{fig:solar day validation} we compare the result for day-time neutrino disappearance $\nu_e\to\nu_e$ probability produced in Sun at $\rho=100$ g/cc and $Y_e=\frac23$ with the well known analytic approximation for two flavors \cite{Parke:1986jy} with some three-flavor information \cite{Fogli:1996ne}:
\begin{align}
P^\odot(\nu_e\to\nu_e) &\approx\frac1{2}\left[1+\cos2\theta_{12} \cos2\theta^\odot_{12}\right] c_{13}^4 +s_{13}^4, \label{eq:PeeD} \\
\text{with} \quad \quad\quad & \notag \\
& \hspace*{-2cm} \cos2\theta^\odot_{12} \equiv \frac{\cos2\theta_{12}-c_{13}^2A_{mat}/\Delta m^2_{21}}{\sqrt{(\cos2\theta_{12}-c_{13}^2A_{mat}/\Delta m^2_{21})^2+\sin^22\theta_{12}}} \,. \notag
\end{align}
The superscript $^\odot$,  indicates the quantity is evaluated at the production point in the sun.

The small difference at high energies is due to the beginning of $\nu_3$ matter effects which are not encompassed in eq.~\ref{eq:PeeD}.
To leading order, these effects can be included by replacing
\begin{align}
    c^4_{13} &\rightarrow  c^2_{13} \cos^2 \theta^\odot_{13} \approx c^4_{13}(1-2s^2_{13} A_{mat}/\Delta m^2_{ee})\,, \notag \\
     s^4_{13} &\rightarrow s^2_{13} \sin^2 \theta^\odot_{13} \approx s^4_{13}(1+2c^2_{13} A_{mat}/\Delta m^2_{ee})\,,
\end{align}
in eq.~\ref{eq:PeeD}, see \cite{Khan:2019doq}.
With this correction, we have numerically confirmed that the difference from the full expression is smaller than the width of the lines in Fig.~\ref{fig:solar day validation} at the highest energy.

\begin{figure}
\centering
\includegraphics[width=\columnwidth]{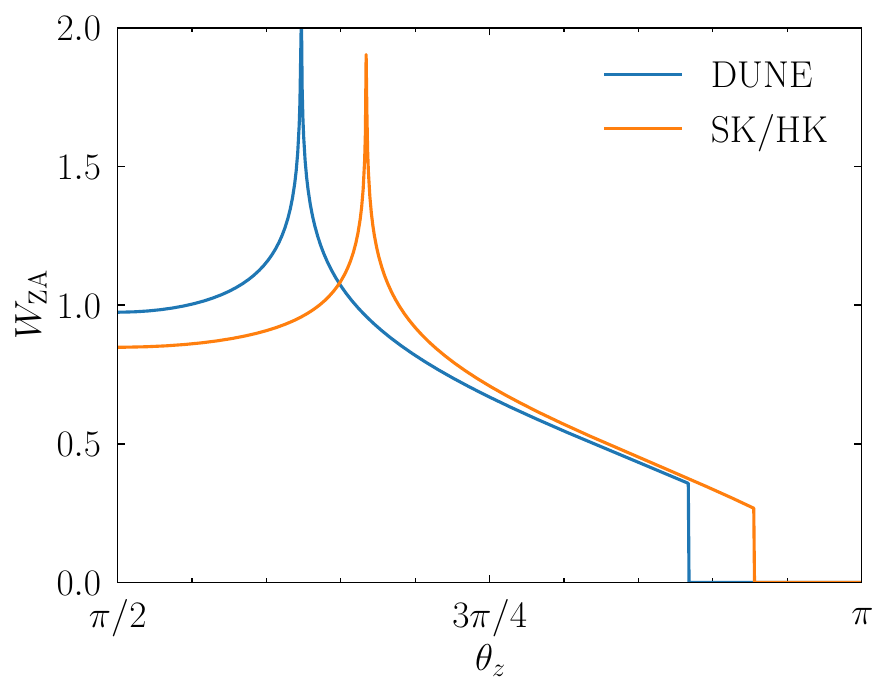}
\caption{The average weight of night-time exposure over a year of solar neutrino measurements at the latitudes of DUNE, 44$^\circ$~N (blue) and SK/HK, 36$^\circ$~N (orange).}
\label{fig:solar weight validation}
\end{figure}

In Fig.~\ref{fig:solar weight validation} we show the average nighttime weight of zenith angles, $W_{\rm ZA}$, an experiment sees over the course of a year at night, see \cite{Lisi:1997yc}.
The weight is defined such that the averaged probability is
\begin{equation}
\langle P^\odot_{ee}\rangle(E)=\int d\theta_z W_{\rm ZA}(\theta_z)P^\odot_{ee}(E,\theta_z)\,.
\end{equation}
If one instead prefers to integrate in $d(\cos\theta_z)$ space, one needs to multiply by the Jacobian, $1/\sin\theta_z$.

\begin{figure*}
\centering
\includegraphics[width=0.49\linewidth]{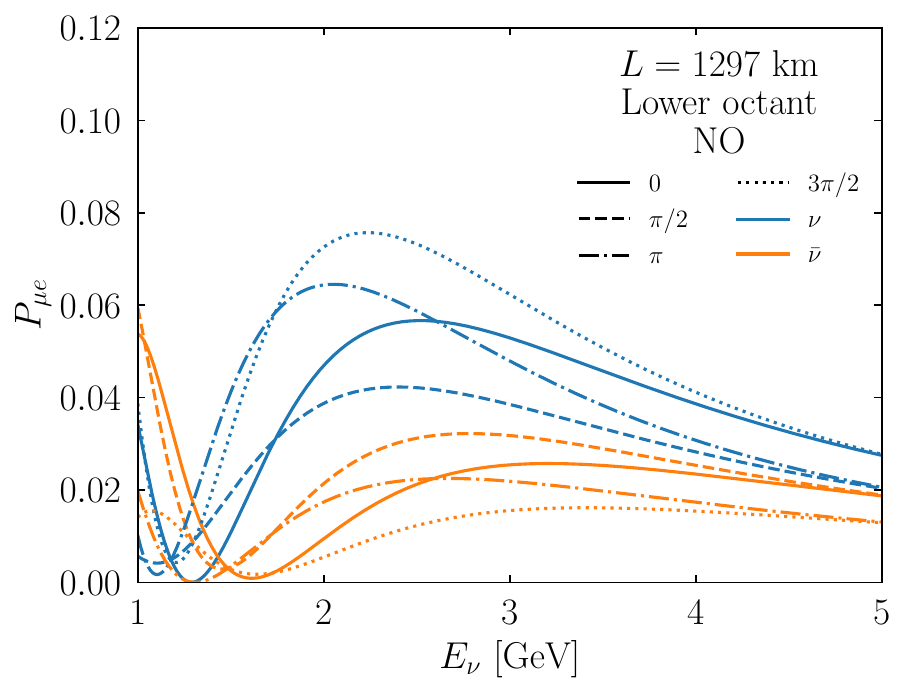}
\includegraphics[width=0.49\linewidth]{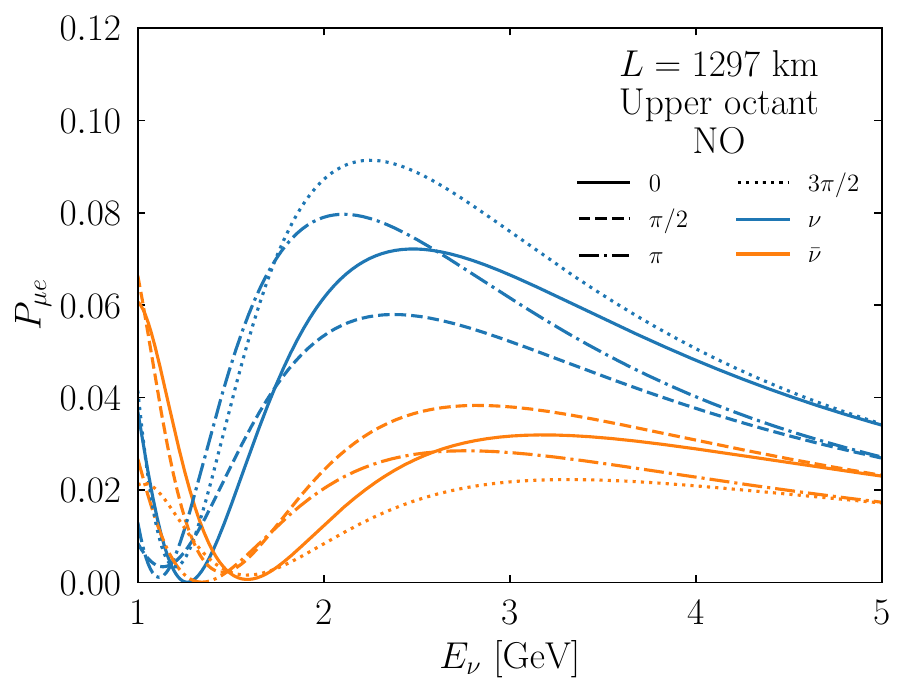}
\includegraphics[width=0.49\linewidth]{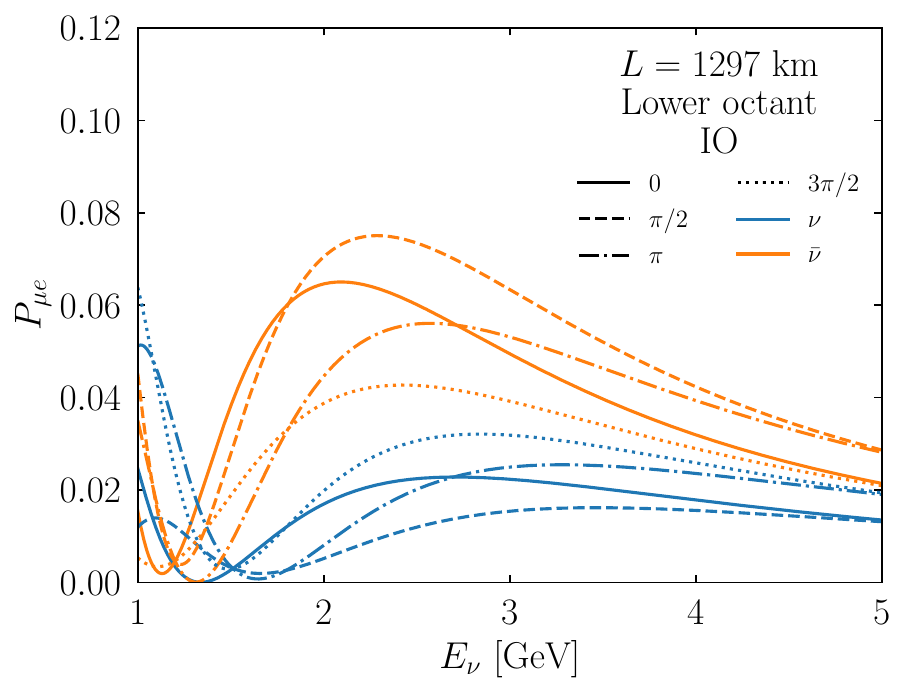}
\includegraphics[width=0.49\linewidth]{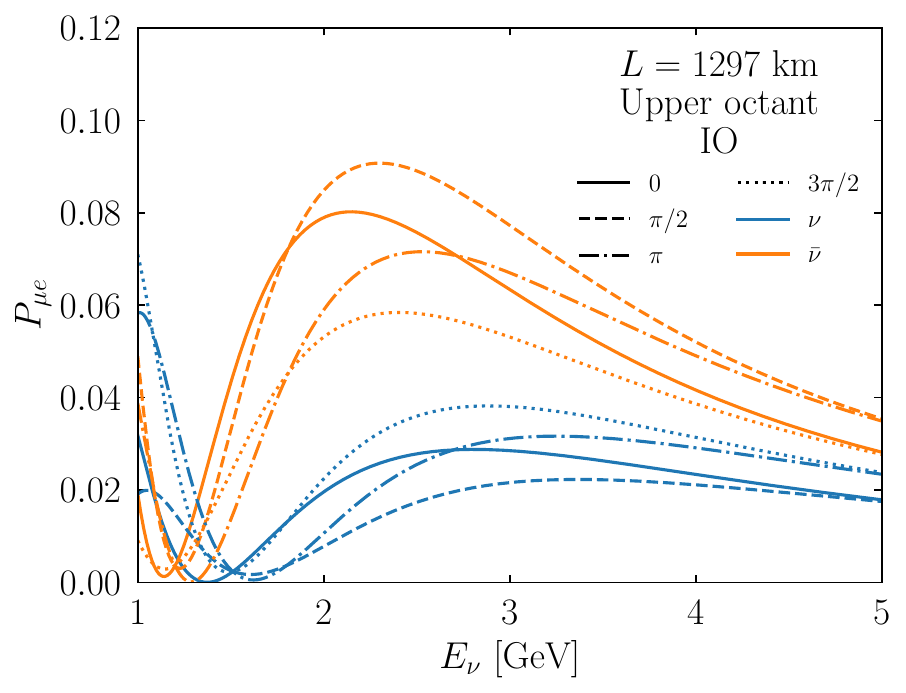}
\caption{The appearance probability for DUNE-like parameters, for both neutrino (blue) and antineutrino (orange) mode and for different values of $\delta$ (denoted by different line styles); the curves follow in a comparable fashion for the shorter NOvA and T2K/HK baselines.
The \textbf{top} panels are for the normal atmospheric mass ordering and the \textbf{bottom} panels are for the inverted ordering.
The \textbf{left} panels are for the lower octant of $\theta_{23}$ ($\sin^2 \theta_{23} = 0.44$), and the \textbf{right} panels are for the upper octant, ($\sin^2 \theta_{23} = 0.56$).}
\label{fig:LBL appearance validation}
\end{figure*}

In Fig.~\ref{fig:LBL appearance validation} we compare to standard long-baseline appearance results in an Earth model that is constant density at $\rho=3$ g/cc and at a zenith angle that corresponds to $L=1297$ km with the source and detector on the surface, roughly the set up for DUNE.
The $\nu_\mu\to\nu_e$ channel is in the \texttt{[1][0]} element of the matrix in the code.
We see the expected variation of the oscillation maximum as $\delta$, the octant of $\theta_{23}$, the mass ordering, and $\nu/\bar\nu$ change.

\clearpage

\bibliography{main}

\end{document}